\documentclass{SciPost}

\usepackage{graphicx,amsfonts,amssymb,amsmath,hyperref,dsfont}
\usepackage{placeins}

\newif\ifhyper
\hypertrue
\ifhyper
\hypersetup{
   citecolor = {green},
   colorlinks = {true}, % false
   urlcolor = {blue} % magenta
}
\fi

\newcommand{\textscript}[1]{\textnormal{\scriptsize{#1}}}
\newcommand{\aro}[1]{a_{#1}^{\phantom{\dagger }}}
\newcommand{\ard}[1]{a_{#1}^{\dagger} }
\newcommand{\bro}[1]{b_{#1}^{\phantom{\dagger}} } 
\newcommand{\brd}[1]{b_{#1}^{\dagger} } 

\newcommand{\ket} [1] {\vert #1 \rangle}
\newcommand{\bra} [1] {\langle #1 \vert}
\newcommand{\braket}[2]{\langle #1 | #2 \rangle}

 \newcommand{\aoo}[1]{a_{\mathbf{#1}}^{\phantom{\dagger }}}
 \newcommand{\ad}[1]{a_{\mathbf{#1}}^{\dagger} } 
\newcommand{\bo}[1]{b_{\mathbf{#1}}^{\phantom{\dagger}} } 
\newcommand{\bd}[1]{b_{\mathbf{#1}}^{\dagger} }

 \newcommand{\alo}[1]{\alpha_{\mathbf{#1}}^{\phantom{\dagger }}}
 \newcommand{\ald}[1]{\alpha_{\mathbf{#1}}^{\dagger} } 
\newcommand{\beo}[1]{\beta_{\mathbf{#1}}^{\phantom{\dagger}} } 
\newcommand{\bed}[1]{\beta_{\mathbf{#1}}^{\dagger} } 
\newcommand{\dG}[1]{\delta_G \left({\mathbf{#1}}\right) } 
\newcommand{\coeff}[2]{C_{#1}\left( \bf {#2}  \right)} 
\newcommand{\coeffgamma}[1]{\Gamma\left( \bf {#1}  \right)} 
\newcommand{\uminus}{\scalebox{0.5}[1.0]{\( - \)}}

\def\bra#1{\langle#1\vert}
\def\ket#1{\vert#1\rangle}

\def\Longarrow{\protect\@lra}
\def\@lra{\relbar\joinrel\relbar\joinrel\relbar\joinrel%
          \relbar\joinrel\rightarrow}

\def\Im {\mbox{Im}}
\def\be{\begin{equation}}       \def\ee{\end{equation}}
\def\bea{\begin{eqnarray}}      \def\eea{\end{eqnarray}}
\def\bes{\begin{subequations}}  \def\ees{\end{subequations}}

\def\k{\mathbf{k}}
\def\Q{\mathbf{Q}}

\newcommand{\w}[1]{\omega\left({\bf #1}\right)} 
\newcommand{\Coeff}[5]{C_{#1}\left(\bf{#2},\bf{#3},\bf{#4},\bf{#5}\right)} 

\newcommand{\pp}{\\[0.1cm]}

\newcommand{\szI}[3]{s^{z}_{1} \left(\bf{#1}, \bf{#2}, \bf{#3}  \right)}
\newcommand{\szII}[3]{s^{z}_{2} \left(\bf{#1}, \bf{#2}, \bf{#3}  \right)}
\newcommand{\szIII}[3]{s^{z}_{3} \left(\bf{#1}, \bf{#2}, \bf{#3}  \right)}
\newcommand{\szIV}[3]{s^{z}_{4} \left(\bf{#1}, \bf{#2}, \bf{#3} \right)}
\newcommand{\spI}[1]{s^{+}_{1} \left(\bf{#1}  \right)}
\newcommand{\spII}[1]{s^{+}_{2} \left(\bf{#1} \right)}
\newcommand{\spIII}[4]{s^{+}_{3} \left(\bf{#1}, \bf{#2} , \bf{#3} , \bf{#4}\right)}
\newcommand{\spIV}[4]{s^{+}_{4} \left(\bf{#1}, \bf{#2} , \bf{#3} , \bf{#4}\right)}
\newcommand{\spV}[4]{s^{+}_{5} \left(\bf{#1}, \bf{#2} , \bf{#3} , \bf{#4}\right)}
\newcommand{\spVI}[4]{s^{+}_{6} \left(\bf{#1}, \bf{#2} , \bf{#3} , \bf{#4}\right)}
\newcommand{\spVII}[4]{s^{+}_{7} \left(\bf{#1}, \bf{#2} , \bf{#3} , \bf{#4}\right)}
\newcommand{\spVIII}[4]{s^{+}_{8} \left(\bf{#1}, \bf{#2} , \bf{#3} , \bf{#4}\right)}
\newcommand{\smI}[1]{s^{-}_{1} \left(\bf{#1} \right)}
\newcommand{\smII}[1]{s^{-}_{2} \left(\bf{#1} \right)}
\newcommand{\smIII}[4]{s^{-}_{3} \left(\bf{#1}, \bf{#2} , \bf{#3} , \bf{#4}\right)}
\newcommand{\smIV}[4]{s^{-}_{4} \left(\bf{#1}, \bf{#2} , \bf{#3} , \bf{#4}\right)}
\newcommand{\smV}[4]{s^{-}_{5} \left(\bf{#1}, \bf{#2} , \bf{#3} , \bf{#4}\right)}
\newcommand{\smVI}[4]{s^{-}_{6} \left(\bf{#1}, \bf{#2} , \bf{#3} , \bf{#4}\right)}
\newcommand{\smVII}[4]{s^{-}_{7} \left(\bf{#1}, \bf{#2} , \bf{#3} , \bf{#4}\right)}
\newcommand{\smVIII}[4]{s^{-}_{8} \left(\bf{#1}, \bf{#2} , \bf{#3} , \bf{#4}\right)}

\newcommand{\nn}{\nonumber \\}
\newcommand{\nnp}{\nonumber \\[0.1cm]}

\def\non{\nonumber}

\def\k{{\bf k}}

\begin{document}

\begin{center}{\Large \textbf{
Mutually attracting spin waves in the square-lattice quantum antiferromagnet
}}\end{center}

\begin{center}
M. Powalski\textsuperscript{1}, 
K.P. Schmidt\textsuperscript{2$^\diamond$}, 
G.S. Uhrig\textsuperscript{1*}
\end{center}

\begin{center}
{\bf 1} Lehrstuhl f\"{u}r Theoretische Physik 1, TU Dortmund, Germany
\\
{\bf 2} Lehrstuhl f\"{u}r Theoretische Physik I, Staudtstra\ss{}e 7, Universit\"at Erlangen-N\"urnberg, D-91058 Erlangen, Germany
\\
$^\diamond$ kai.phillip.schmidt@fau.de\\
* goetz.uhrig@tu-dortmund.de
\end{center}

\begin{center}
\today
\end{center}

\section*{Abstract}
{\bf 
Spin waves (magnons) in two dimensions are the potential glue in 
high-temperature superconductors so that their quantitative understanding 
is mandatory. Yet even for the fundamental case of the undoped Heisenberg 
model on the square lattice a consistent picture is still lacking. 
Significant spectral continua are taken as evidence of the existence 
of fractional excitations (spinons), but descriptions
in terms of spinons fail to show the established absence of an energy
gap. Here a fully consistent picture of the dynamics in the
square-lattice quantum antiferromagnet is provided which agrees with 
the experimental findings. The key step is to capture (i) the strong 
attractive interaction between the spin waves and (ii) 
the vertex corrections of the observables.
}

\vspace{10pt}
\noindent\rule{\textwidth}{1pt}
\tableofcontents\thispagestyle{fancy}
\noindent\rule{\textwidth}{1pt}
\vspace{10pt}

\section{Introduction}

The spin-$\frac{1}{2}$ Heisenberg model is one of the simplest and most 
paradigmatic models in quantum magnetism \cite{harri67}.
Its relevance has been boosted further by the discovery of cuprate
high-temperature ($T_c$) superconductors \cite{bedno86} as their undoped 
parent compounds realize the Heisenberg model on the square-lattice 
\cite{manou91}. Tremendous experimental and theoretical effort has 
been invested in studying their magnetic excitations 
which are believed to provide the glue of high-$T_c$ superconductivity. 

Much is known about the elementary excitations at large wavelengths 
\cite{reger88,chakr89,auerb91,sandv97b}, described by spin waves (magnons), 
the Goldstone bosons of the long-range ordered antiferromagnetic phase 
\cite{golds62}.  But their nature at short wavelengths
 remains unclear to this day. 
Yet  precisely the short-range processes  play 
a decisive role in the understanding of high $T_c$ superconductivity 
\cite{vigno07,lipsc07,dahm09,letac11}.

\begin{figure*}
\begin{center}
\includegraphics[scale=0.4]{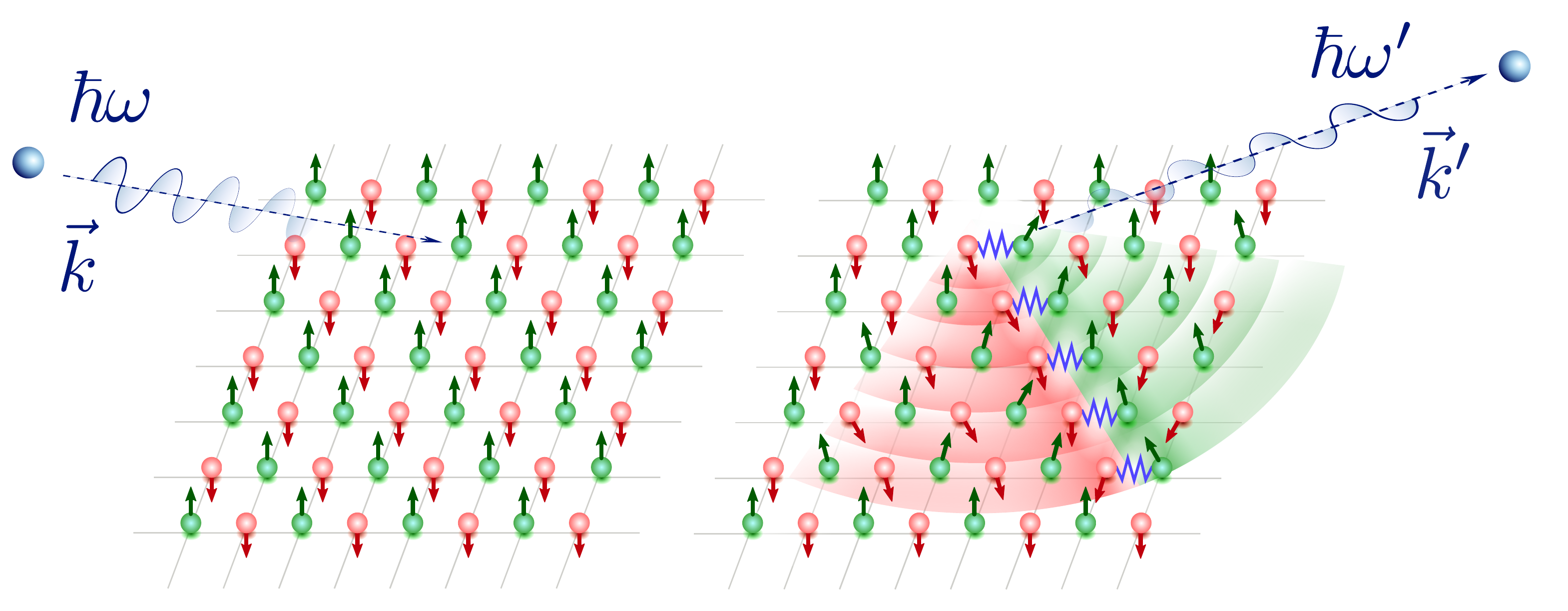}  
\end{center}
\caption{ \label{fig:sketch}
\textbf{Sketch of a scattering event}
By inelastic scattering of a neutron in the longitudinal channel
two magnons are excited from the long-range ordered antiferromagnetic ground state.
On short distances, they interact strongly and attract each other
forming resonances \cite{powal15}. Note that the magnons are sketched here as wave packets
localized in \emph{real} and in \emph{reciprocal} space with a finite minimum uncertainty
as required by Heisenberg's uncertainty principle.}
\end{figure*}

Since the seminal paper by Anderson \cite{ander87} there
are  wide-spread activities to seek for fractionalization of 
magnons into spinons \cite{ho01,chris07a,dalla15}.
By contrast, here we present strong evidence that spinons do not appear as 
the elementary excitations at any wavevector. 
We derive a comprehensive picture in terms of dressed magnons which agrees 
strikingly with experimental data.
Their anomalous behavior at large wavevectors can be traced back to a strong 
mutual attraction 
on short-length scales as sketched in Fig.\ \ref{fig:sketch}. Thus, our 
results provide 
key information on the nature of the magnetic excitations in the 
important class of parent compounds of high-$T_c$ superconductors.

The quantum antiferromagnet on the square lattice (QASQ) is a paradigmatic 
example of long-range ordered quantum phases and broken continuous symmetries 
in condensed matter~\cite{manou91,braun10}. 
In a previous work \cite{powal15}, we derived a quantitative description of the 
magnon  dispersion including the roton minimum at high energies.
The key step was to take the full renormalization of
the magnon dispersion and of the magnon-magnon interactions into account. 
Technically, this was achieved by a continuous similarity transformation (CST)
which was performed in a non-perturbative fashion, see also Sect.\ \ref{sec3}.
This means that we changed the basis in which the model is described
in such a way that the resulting effective model is easier to analyze.
The attractive magnon-magnon interaction gives rise to the formation of a 
two-magnon ``Higgs'' resonance corresponding to a longitudinal magnon with
finite lifetime. These calculations in the longitudinal channel were based on 
the bare observable, i.e., the observable was not transformed to the  basis
used for the Hamiltonon operator. This means that the vertex corrections
were still lacking completely. Yet, they are absolutely crucial for
the continuum in the transversal spin channel. Without them, no such
continuum occurs. This vital piece of understanding of 
the inelastic scattering of polarized neutrons is added in the
present work.

We aim at the quantitative description of static and 
dynamic correlations of single- and multiple-magnon states. This will 
allow us to compare the theoretical spectroscopic signatures of interacting 
magnons with recent experimental data.
For this purpose we derive the effective observables which embody the vertex corrections.   
The systematic consideration of effective observables is a crucial extension 
of Ref.\ \cite{powal15} because it allows us to analyze the spectroscopic 
features quantitatively. We emphasize that our approach does not 
require any resort to fractional spinons.

%\onecolumngrid
%Figure 1: Nice Plot
%%%%%%%%%%%%%%%%%%%
%%%%%%%%%%%%%%%%%%%
%\twocolumngrid

We show that the anomalous dispersion in the magnon spectrum is caused by
substantial hybridization of single magnons with the three-magnon continuum. 
This hybridization is strongly enhanced by a strong attraction between pairs of 
magnons leading to significant continua. Our interpretation is supported by the
noticeable agreement with recent experimental results. 

The paper is organized as follows.
In Sect.\ \ref{sec2}, we discuss the major intricacies which arise in the general magnon approach pointing out the relevance of magnon-magnon interactions.
Section \ref{sec3} renders a brief overview of the technical approach 
based on continuous transformations in momentum space. 
In Sect.\ \ref{sec4}, we present the spectroscopic properties of interacting
magnons and compare the theoretical results with experimental data from inelastic scattering with polarized neutrons.
Our results are interpreted  theoretically in Sect.\ \ref{sec5}. 
We identify the origin of the increased weight in the 
high-energy spin wave continuum and discuss the relevance of mutual magnon attraction. The final Sect.\  \ref{sec6} provides the conclusions and an outlook.

%%%%%%%%%%%%%%  SECTION II  %%%%%%%%%%%%%%%
%%%%%%%%%%%%%%%%%%%%%%%%%%%%%%%%%%%%%%%%%%%
\section{Renormalized magnon description}
\label{sec2}
%%%%%%%%%%%%%%%%%%%%%%%%%%%%%%%%%%%%%%%%%%%
%%%%%%%%%%%%%%%%%%%%%%%%%%%%%%%%%%%%%%%%%%%

There is no exact solution of the  QASQ, but its low-energy physics is well understood. There is overwhelming evidence that the ground state exhibits 
long-range N\'eel order for zero temperature~\cite{manou91}. This ordering is associated with a finite staggered magnetization which spontaneously breaks the
continuous SU(2) symmetry of the  Hamiltonian.
As a result, the QASQ displays gapless bosonic excitations in accordance with 
Goldstone's theorem~\cite{golds62}. The corresponding bosons are quantized 
spin waves, i.e., magnons, which exhibit a linear dispersion at 
long wavelengths~\cite{ander52,kubo52}. 
Even in the case $S=1/2$, where quantum fluctuations are strongest \cite{auerb91},
magnons provide a quantitative and consistent description of the low-energy excitations. But their nature at high energies, i.e., at short wavelengths, 
is still not clarified to this day.

In the following, we give a brief sketch of the general spin wave formalism for 
the QASQ focusing on the physical interpretation and the intricacies of this approach.  Details can be found in various standard textbooks 
\cite{matti88, auerb94}, reviews \cite{manou91,katan07},
 or in the original papers \cite{ander52,kubo52,dyson56a,malee58,oguch60,zitta65}.
We consider the Heisenberg model with nearest-neighbor antiferromagnetic
exchange interaction $J>0$ between spins with $S=1/2$ 
%Starting Hamiltonian
%%%%%%%%%%%%%%%%%%%%%%%
\be
 H = J \sum_{\langle i,j\rangle} \vec{S}_{i}\cdot \vec{S}_{j}\quad.
 \label{eq:hamiltonian}
\ee
%%%%%%%%%%%%%%%%%%%%%%%
%
At zero temperature the ground state exhibits long-range N\'eel order 
implying that the spin orientations favor an anti-parallel alignment on 
sublattice A and B of the bipartite square lattice. 
The classical N\'eel state $\ket{\text{AF}}$ with 
$\uparrow$ spins on A and $\downarrow$ spins on B 
is chosen as a reference state of the antiferromagnetic system.
The main idea of spin wave theory is to represent the deviations from this
 classical N\'eel state by bosonic excitations.
To this end, spin operators are expressed by local bosonic creation and 
annihilation operators $a^{(\dagger)}_i$ and $b^{(\dagger)}_j$ discriminating between bosons on sublattice $A$ and $B$.

Here we choose the Dyson-Maleev representation \cite{dyson56a,malee58,katan07}
defined by the following relations
\bes
\label{eq:dyson_maleev_representation}
\bea
S_i^{+} &=& \sqrt{2S} \left(\aro{i} - \frac{\ard{i} \aro{i} \aro{i}}{2S}\right) \qquad  
S_i^{\uminus} =  \sqrt{2S}\, \ard{i}  
 \\
S_j^{+} &=& \sqrt{2S} \left(\brd{j} - \frac{\brd{j} \brd{j} \bro{j}}{2S}\right) 
\qquad \; S_j^{\uminus} =  \sqrt{2S}\,\bro{j} 
\\
S^z_i &=& S - \ard{i}\aro{i} \qquad S^z_j = -S + \brd{j}\bro{j}\quad.  
\eea
\ees
The resulting Hamiltonian is equivalent to coupled harmonic oscillators 
with additional anharmonic interactions. 
The  essential advantage of the Dyson-Maleev representation
is that the interactions can be expressed by a finite
number of \emph{quartic} boson operators at the expense of the 
manifest hermitecity of the Hamiltonian. By contrast, a
Holstein-Primakoff representation generally requires an
expansion in $1/S$ due to the occurring square root expressions
\cite{holst40,auerb91}.

We stress that the Dyson-Maleev representation is a faithful
representation of the spin algebra. This means that the dynamical
processes expressed by functions of the spins are faithfully expressed
by the bosonic expressions \eqref{eq:dyson_maleev_representation} if they
start from a physical state and end at a physical state, i.e., states
with at maximum $2S$ bosons per site. In the literature 
\cite{dyson56a,zitta65,katan07} the distinction between dynamical and kinematical
interactions is made. The dynamical part is the one which expresses multi-particle
processes (quartic and possibly higher terms in the boson operators) in the Hamiltonian. The kinematical part results in the condition to include only 
physical states, i.e., states with at maximum $2S$ spins per site. We find this
 distinction slightly misleading
because the faithful representation of the spin algebra ensures already that
unphysical states are not reached from physical states or do not lead to physical states.
Thus, the essential aspects are covered by the ``dynamical'' interaction.
This is the basis for the rigorous perturbative analyses of the spin dispersion
and spin correlations \cite{hamer92,igara05,syrom10,majum10,majum12a}.
Only at finite temperatures, the kinematic constraint has to be imposed
additionally because otherwise unphysical states would contribute to the
density operators.

Since we focus here on zero temperature response functions
we henceforth only need  the quartic terms ensuing from the Dyson-Maleev representation 
\eqref{eq:dyson_maleev_representation} and we will call them interactions
if they consist of two incoming and two outgoing bosons. Otherwise, we call them
hybridizations because they link one boson to three bosons. The attribute ``dynamical''
is omitted because the spin algebra ensures the kinematical aspects as well
 at zero temperature.

The bilinear bosonic terms of $H$ can be diagonalized by a Bogoliubov 
transformation in momentum space which introduces 
transformed  bosons $\ald{k} = l_{\mathbf{k}} \ad{k} + m_{\mathbf{k}}\beo{-k}$ and 
$\bed{k} = m_{\mathbf{k}} \aoo{-k} + l_{\mathbf{k}}\bd{k}$ where we used the
 Fourier transformed boson operators 
\be
\ad{k} = \frac{1}{\sqrt{N}} \sum_{\textbf{r}_i\in A} \ard{i} 
\thinspace e^{\textbf{k} \cdot \textbf{r}_{j} } \qquad
\bd{k} = \frac{1}{\sqrt{N}} \sum_{\textbf{r}_j\in B} \brd{j} 
\thinspace e^{\textbf{k} \cdot \textbf{r}_{j} }
\ee 
with $N$ denoting the number of sites of the $A$ \emph{or} the $B$ sublattice.
We are working throughout this article with operators in the Hamiltonian
 in the magnetic Brillouin zone (MBZ) excluding 
the region around $\mathbf{k}=(\pi,\pi)$. 
Hence, the two gapless modes are found only at the origin while the
dispersion, see below, is finite at the magnetic zone boundary. Note, however, 
that operators appearing in the structure factors and 
representing external (de)excitations can have momenta in the full Brillouin zone.

The explicit coefficients $l_{\mathbf{k}}$ and $m_{\mathbf{k}}$ are given in 
Ref.\ \cite{uhrig13}. The resulting spin wave Hamiltonian 
can be written in the following form
\be
\label{eq:initial_sw_hamiltonian}
H_{\rm init} = H_{\text{SW}} + V^{0} + V^{+} + V^{-} \quad,
\ee
where the bilinear part 
\be
H_{\rm SW} = E_0 + \sum_{\k} \omega_{\k} \left(\alpha_{\k}^{\dagger}
\alpha_{\k}^{\phantom{\dagger}}+\beta_{\k}^{\dagger}
\beta_{\k}^{\phantom{\dagger}}\right)  
\ee
describes decoupled magnons modes. The coefficients 
\bes
\bea
e_0^{(1)} & := & E_0/N =-2J (S^2+AS +A^2/4)
\\
\omega_{\mathbf{k}} & =& 2J(2S+A)\sqrt{1-\gamma
\left(\mathbf{k}\right)^2}
\\
& =:& (1+A/(2S)) \omega_{\mathbf{k}}^{(0)}
\\
A & := & \frac{2}{N}\sum_{\mathbf{k}}\left(1-\sqrt{1-\gamma\left(\mathbf{k}\right)^2}
\right) \approx 0.157947
\eea
\ees
correspond to the ground-state energy per site and the one-magnon dispersion in next-to-leading order spin wave theory~\cite{hamer92,uhrig13}.

The remaining non-diagonal part can be decomposed into quartic
interaction terms
\be
V^{0} = \sum_{\mathbf{1234}}
\delta^{\mathbf{1}\mathbf{2}}_{\mathbf{3}\mathbf{4}}
\Big\{
V_{\mathbf{1}\mathbf{2}\mathbf{3}\mathbf{4}}^{(\alpha\alpha)} \,
\ald{1}\ald{2} \alo{3}\alo{4} 
+V_{\mathbf{1}\mathbf{2}\mathbf{3}\mathbf{4}}^{(\beta\beta)} \,
\bed{\uminus4}\bed{\uminus3} \beo{\uminus2}\beo{\uminus1} +
V_{\mathbf{1}\mathbf{\uminus2}\mathbf{3}\mathbf{\uminus4}}^{(\alpha\beta)} \,
\ald{1}\alo{3} \bed{\uminus 4} \beo{\uminus 2} \Big\}
\label{eq:interact}
\ee
which conserve the number of magnons and into quartic
hybridization processes
\bes
\label{eq:hybrid}
\bea
V^{+} &=& \sum_{\mathbf{1234}}
\delta^{\mathbf{1}\mathbf{2}}_{\mathbf{3}\mathbf{4}}
\Big\{
V_{\mathbf{1}\mathbf{2}\mathbf{3}\mathbf{4}}^{(3 \uminus \alpha)} \,
\ald{1}\ald{2} \bed{\uminus3}\alo{4} 
+V_{\mathbf{1}\mathbf{2}\mathbf{3}\mathbf{4}}^{(3\uminus\beta)} \,
\ald{1}\bed{\uminus3}\bed{\uminus 4}\beo{\uminus2} +
V_{\mathbf{1}\mathbf{2}\mathbf{3}\mathbf{4}}^{(+4)} \,
\ald{1}\ald{2} \bed{\uminus3} \bed{\uminus4} \Big\}
\\
V^{-} &=& \sum_{\mathbf{1234}}
\delta^{\mathbf{1}\mathbf{2}}_{\mathbf{3}\mathbf{4}}
\Big\{
V_{\mathbf{1}\mathbf{2}\mathbf{3}\mathbf{4}}^{(\alpha \uminus 3)} \,
\ald{1}\alo{3}\alo{4}\beo{\uminus2} 
+V_{\mathbf{1}\mathbf{2}\mathbf{3}\mathbf{4}}^{(\beta\uminus 3)} \,
\alo{3}\bed{\uminus 4}\beo{\uminus 1}\beo{\uminus2} +
V_{\mathbf{1}\mathbf{2}\mathbf{3}\mathbf{4}}^{(-4)} \,
\alo{3}\alo{4} \beo{\uminus1} \beo{\uminus2} \Big\}
\eea
\ees
which increase~($+$) or decrease~($-$) the number of magnons. 
Note that the changes of the number of magnons are \emph{even}
as a consequence of the collinear N\'eel order and the bipartiteness
of the square lattice. Either two or four magnons are created in 
$V^+$ or annihilated in $V^-$.
We use a shorthand notation for the momenta $\mathbf{k}_i \rightarrow 
\mathbf{i}$ such
 that $a^{\dagger}_{\mathbf{k}_i} := \ad{i}$ to lighten the notation. The 
Kronecker symbol $\delta^{\mathbf{1}\mathbf{2}}_{\mathbf{3}\mathbf{4}}$ equals
unity if $\mathbf{1}+\mathbf{2}=\mathbf{3}+\mathbf{4}$ modulo a 
reciprocal lattice vector and zero 
otherwise ensuring the conservation of total crystal momentum. 
The explicit vertex functions are given in Ref.\ \cite{uhrig13}.

The interaction and hybridization processes are a consequence of  
the algebraic properties of the spin operators represented by boson operators.
They include the constraint of finite dimensional local Hilbert spaces
if processes starting and ending at physical states are considered.
The hybridization of single magnons with three-magnon states
expressed in $V^{+}$ and $V^{-}$ turns the system into an intricate 
many-body problem. The interaction and the hybridization terms 
are proportional to $1/S$ as is obvious from the Dyson-Maleev 
representation \eqref{eq:dyson_maleev_representation}. Thus, on the one hand, 
it is to be expected that their effect is particularly strong for $S=1/2$. 
On the other hand, one finds that long-wavelength magnons propagate almost freely because the deviation from the N\'eel state is distributed in real space 
and the scattering due to the interaction is relatively small 
\cite{dyson56a,oguch60,zitta65}. This is derived by mapping the 
microscopic model in the long-wavelength limit to a continuum model and analyzing
it by renormalization group techniques \cite{chakr89}.
Thus, the effect of the interaction terms is marginal in the limit of low energies and long wavelengths which is also
 reflected by the good accuracy of the conventional 
next-to-leading order spin wave theory where one neglects the terms 
${V}^{\pm}$ and ${V}^{0}$. This approximation is essentially based on the 
assumption that $\langle a^{\dagger}_i a^{\phantom{\dagger}}_i 
\rangle,  \langle b^{\dagger}_j b^{\phantom{\dagger}}_j \rangle \ll S$. 

But in the case of magnons at short wavelengths, the interactions 
\eqref{eq:interact} as well as the hybridizing terms \eqref{eq:hybrid}
can play a role in spite of their short range in real space.
The short-wavelength magnons have high energies and thus their
scattering and hybridizing processes dispose of a much larger phase space.
A simplistic visualization of this fact is that magnons
at short wavelengths can form much better localized wave packets
at given relative uncertainty $\Delta k/k$ than magnons at
long wavelengths.
As a result, the conventional perturbative expansion in powers 
of $1/S$ turns out to be inefficient displaying only a slow convergence
 at $\mathbf{k}=(\pi,0)$ as reported in Ref.~\cite{syrom10}. 
We showed that the anomalous dispersion of the high-energy magnons at the zone boundary of the MBZ can be attributed to strong quantum effects caused by the interplay of magnon attraction and the hybridization terms at short wavelengths \cite{powal15}.
Consequently, the deviations of spin wave theory at the zone boundary are rather 
a result of a methodical insufficiency than indicating fundamentally 
different physics. The non-perturbative nature of high-energy magnons requires 
a sophisticated methodical treatment which goes beyond the 
conventional perturbative expansion in $1/S$.

\subsection{Effective magnonic Hamilton operator}

The fundamental idea of our approach is to map the initial Hamilton operator
to an effective  Hamilton operator in terms of magnons
\be
\label{eq:eff_sw_hamiltonian}
\mathcal{H}_{\text{eff}} = \mathcal{H}_{\rm M}+ \mathcal{V}^{0}
\ee
which conserves the number of magnons because no
hybridization terms $\mathcal{V}^{\pm}$ appear anymore. 
The calligraphic $\mathcal{V}$ is used for the \emph{renormalized} 
interaction after the change of basis. The initial quartic terms
were denoted by straight $V$. The first part
\be
\label{eq:eff_single_spinwaves}
\mathcal{H}_{\text{M}} = \mathcal{E}_0 +  \sum_{\mathbf{k}} 
\omega_{\mathbf{k}}^{\text{eff}} 
\left(\alpha_{\k}^{\dagger}\alpha_{\k}^{\phantom{\dagger}}+\beta_{\k}^{\dagger}\beta_{\k}^{\phantom{\dagger}}\right) 
\ee
describes renormalized magnons. 
The second part denotes the effective interactions 
\be
\label{eq:eff_interactions}
\mathcal{V}^{0} = \sum_{\mathbf{1234}}
\delta^{\mathbf{1}\mathbf{2}}_{\mathbf{3}\mathbf{4}}
\Big\{\mathcal{V}_{\mathbf{1}\mathbf{2}\mathbf{3}\mathbf{4}}^{(\alpha\alpha)} \,
\ald{1}\ald{2} \alo{3}\alo{4} 
+\mathcal{V}_{\mathbf{1}\mathbf{2}\mathbf{3}\mathbf{4}}^{(\beta\beta)} \,
\bed{\uminus4}\bed{\uminus3} \beo{\uminus2}\beo{\uminus1} +
\mathcal{V}_{\mathbf{1}\mathbf{2}\mathbf{3}\mathbf{4}}^{(\alpha\beta)} \,
\ald{1}\alo{3} \bed{\uminus 4} \beo{\uminus 2} \Big\}\quad.
\ee
We stress that in the continuous transformation the effect of the 
hybridization processes in \eqref{eq:initial_sw_hamiltonian} are absorbed 
into the renormalized coefficients, i.e., into the effective ground state energy 
$\mathcal{E}_0$, the magnon dispersion $\omega_{\mathbf{k}}^{\text{eff}}$,
and the effective interaction $\mathcal{V}^{0}$.
In this way, the hybridization processes are accounted for by the 
renormalized Hamiltonian $\mathcal{H}_{\text{eff}}$  of the dressed magnons 
which constitute the true elementary excitations of the system.  

\subsection{Spectral densities}

Spectral densities provide the theoretical description of the 
momentum and frequency resolved counting rates 
$I_{\textscript{exp}}(\omega, \mathbf{Q})$ measured in INS experiments. 
For sufficiently low temperatures the rate $I_{\textscript{exp}}(\omega, \mathbf{Q})$  is proportional to the dynamic structure factor (DSF) at zero temperature 
 \be
\label{eq:zz}   
S^{\alpha \alpha}(\omega, \mathbf{Q})  = 
	-\frac{1}{\pi} \Im{\ \bra{0}S^{\alpha}_{\text{eff}}(-\mathbf{Q}) 
		\frac{1}{\omega  - (\mathcal{H}_{\text{eff}}-  \mathcal{E}_0)} 
	S^{\alpha}_{\text{eff}}(\mathbf{Q})\ket{0}}             
\ee
where $\ket{0}$ is the ground state of the effective Hamiltonian 
$\mathcal{H}_{\text{eff}}$ with the renormalized ground state energy 
$\mathcal{E}_0$ and the $S^{\alpha}_\text{eff}(\mathbf{Q})$ are the 
Fourier transformed  components of the effective spin operator with 
$\alpha=x,y,z$.

Let us assume that the staggered magnetization of the antiferromagnetic phase,
which breaks the spin-rotation symmetry spontaneously,
 is aligned along the $z$-direction.
Inelastic scattering experiments with polarized neutrons allow one to distinguish
between the longitudinal part of the DSF for $\alpha=z$ and its transversal parts at
$\alpha=x,y$ which probe the longitudinal or transversal excitations, respectively.
Since we consider an isotropic Hamiltonian the DSF is invariant under the exchange of the $x$ and $y$ direction. In this case, it is expedient to define the transversal DSF as follows 
\be
\label{eq:xy}
 S^{xx+yy}(\omega, \mathbf{Q}) := S^{xx}(\omega, \mathbf{Q})+S^{yy}(\omega, \mathbf{Q}) = 
-\frac{1}{\pi} \Im{\ \bra{0}S^{-}(-\mathbf{Q}) \frac{1}{\omega  - 
	(\mathcal{H}_{\text{eff}}-  \mathcal{E}_0)} S^{+}(\mathbf{Q})\ket{0} }           
\ee
combining the $x$ and $y$ contribution, see also Ref.\ \cite{canal93}.
From Eqs.\ \eqref{eq:zz} and \eqref{eq:xy} we learn that we have
to compute the corresponding resolvents to determine their
spectral densities.

Due to the conservation of the number of magnons in $\mathcal{H}_{\text{eff}}$ 
the Hilbert subspaces of different numbers of magnons do not interact.
Hence, in subsequent calculations each subsector of $n$ magnons
can be treated separately~\cite{knett03a,powal15}. This facilitates the computations greatly
because it converts a true many-body problem into a few-body problem.
For instance, the resolvent $R(\omega, \mathbf{Q})$ yielding the spectral densitiy 
$S(\omega, \mathbf{Q})=-\frac{1}{\pi} \Im{R(\omega,\mathbf{Q})}$
splits into contributions from separate subspaces of
different numbers $n$ of magnons 
\be
\label{eq:resolvent}
R(\omega, \mathbf{Q})	= 
\sum_n \bra{0}O^{(-n)}(\mathbf{-Q}) \frac{1}{\omega  - 
(\mathcal{H}_{\text{eff}}-  \mathcal{E}_0)} {O}^{(+n)}(\mathbf{Q})\ket{0}  
\ee
where $O^{(n)}$ stands for the part of the effective observable creating
$n$ magnons (subscript $(+n)$) or annihilating $n$ magnons (subscript $(-n)$).

Since the original Hamilton operator changes the number of
magnons only by pairs the effective observables which have
contributions in the one-magnon sector, will generically
also have contributions in the three-magnon sector and higher.
The effective observables which have
contributions in the two-magnon sector, will generically
also have contributions in the four-magnon sector and higher.
The transversal DSF  $S^{xx+yy}(\omega, \mathbf{Q})$ results from
$O=S^{\pm}_{\text{eff}}(\mathbf{Q})$ and couples to 
the odd sectors $n=1$ and $n=3$. Higher contributions, e.g., $n=5$
are negligible as is supported by sum rules, see below.
In the longitudinal DSF $S^{zz}(\omega, \mathbf{Q})$ 
the even sector $n=2$ dominates which is again supported
by a sum rule, see below.
The associated \emph{static} structure factors (SSF) are given by 
\be
\label{eq:static}
S^{\alpha\alpha}_n(\mathbf{Q}):=
\langle 0|S^\alpha(-\mathbf{Q}) \mathcal{P}_n S^\alpha(\mathbf{Q})|0\rangle
\ee
where $\mathcal{P}_n$ projects onto the subspace with $n$ magnons.
Obviously, the SSFs provide the 
frequency integrals of the spectral densities of $R(\omega,\mathbf{Q})$ in
\eqref{eq:resolvent}.

If the resolvent in a subspace with more than a single magnon is computed, 
the interaction of each pair of magnons must be taken into account. 
Previously, this effect was not accounted
for in spin wave calculations \cite{canal92b,canal93}. The description in
terms of an effective $O(3)$-continuum model with adjusted parameters
includes interaction effects, but is tailored to the Raman response, i.e.,
the response at zero momentum or infinite wavelength \cite{weidi15}.
Hence the results computed for the microscopic lattice model by CST
and subsequent treatment of the remaining few-body problem promote 
our understanding of the QASQ to a significantly higher level 
which has been beyond reach so far.

In the evaluation of the DSFs the dispersion and 
the effective magnon-magnon interaction have to be considered. 
Since we started from the Dyson-Maleev representation of the spin Hamiltonian
\eqref{eq:hamiltonian} the Hamiltonian was not manifestly hermitian.
This remains true during the CST and also after the transformation.
The one-magnon part $\mathcal{H}_\textrm{M}$ is hermitian, but the
interaction part $\mathcal{V}^0$ is not. Thus no standard Lanczos
algorithm is applicable, but a non-symmetric Lanczos algorithm 
\cite{rajak91}
is applied to determine the continued fraction representation of the resolvent.

The magnon operators are represented on a mesh of discrete points
in the MBZ. For the CST, this mesh cannot be
chosen very dense because the number of differential equations
to be solved becomes too large. In order to keep finite-size effects small
in the calculation of spectral densities, we interpolate the coefficients in 
the Hamiltonian and the observables to obtain a denser mesh in momentum space. 
For the longitudinal channel, the system size is enhanced 
in this way from $L=8$ to $L=192$ where $L^2$ defines the number of  points in the MBZ.
For the continua in the transversal channel,  we extrapolated from $L=8$ to $L=16$
which is sufficient in view of the dimension of the Hilbert space
which is $L^4$ in the three-magnon channel because of the two undetermined momenta. 
Still, the discretization needs to be fine enough to capture the continua.

The resulting spectral densities are sums of 
weighted $\delta$-functions because we truncate the depth of the
continued fractions where finite-size effects set it. 
Smooth densities are obtained 
by broadening the truncated continued fractions by replacing the 
$\delta$-peaks by Gaussians with the corresponding weight $W_i$ and a
constant broadening $\sigma$
  \bea
	\label{eq:gaussian_sd}
	I(\omega) =\sum_i W_i \delta(\omega - \omega_i) \rightarrow  \sum_i W_i  
	\frac{1}{\sqrt{2 \sigma^2 \pi}} \textnormal{e}^{ 
	\frac{1}{2}\left(\frac{\omega -\omega_i}{\sigma}\right)^2}\quad.
	 \eea
There are three reasons for introducing this broadening. The first one 
is to account for the finite experimental resolution. The second one is 
to account for the experimental uncertainty in the determination of the continua. 
The measurement of continua
is more challenging than the determination of pronounced peaks because
the continua are more strongly affected by the possible systematic errors
in the subtraction of the backgrounds. The third reason is to mimic
the finite life time of the measured excitations induced by finite 
temperature and/or by imperfections (disorder) in the sample.
Note that the latter two reasons suggest a broadening depending
on the polarization because different polarizations focus on different
physical processes, e.g., pronounced peaks are only discernible
in the transverse response. Thus the transversal and the longitudinal 
response will generically show a different broadening.

The spectral weights depend slightly on the interpolation of the coefficients
resulting from the numerical CST. In particular, the relative weights between
channels of different magnon number are influenced since different interpolation schemes are employed for sectors of different magnon number.
To ensure the correct relative weights, we rescale the resulting 
spectral densities such that the ratios between the weights,
i.e.,  the frequency integrals, are consistent with the
extrapolations of the static structure factors \eqref{eq:static}
which can be extrapolated reliably to the
thermodynamic limit, i.e., to infinite system size. 

\section{Continuous transformation in momentum space}
\label{sec3}

The continuous change of basis is parametrized by a running parameter 
$\ell$ starting at $0$ and terminating the transformation at $\ell=\infty$. 
The reciprocal value of $\ell$ represents an  energy cutoff. 
The flowing Hamiltonian $H(\ell)$ is transformed from the initial Hamiltonian 
$\mathcal{H}(\ell=0) =: H_{\rm init}$ to the final effective one 
$H(\ell \rightarrow \infty) =: \mathcal{H}_{\rm eff}$.
This is achieved by integrating the flow equation
\be
\label{eq:flow_equ}
\partial_\ell \mathcal{H}(\ell) = [\eta(\ell),\mathcal{H}(\ell)]
\ee
using the particle conserving generator 
$\eta(\ell)=\mathcal{V}^+(\ell)-\mathcal{V}^{-}(\ell)$  
\cite{knett00a,fisch10,powal15}. This generator ensures that the effective Hamiltonian exhibits the desired form \eqref{eq:eff_sw_hamiltonian}.  All
terms which net create or annihilate particles are rotated away in the 
course of the flow \eqref{eq:flow_equ} \cite{knett00a,knett03a}.
The resulting differential equations are presented in the Appendix
\ref{app:flowham}.

The relevant observables $O$, i.e., the spin operators $S^{-}(-\mathbf{Q})$, 
$S^{+}(-\mathbf{Q})$ and $S^z(\mathbf{Q})$, are transformed as well.
In the framework of the CST this is achieved by applying the same generator
\be
\label{eq:flow_obs}
\partial_\ell O =[\eta(\ell),O(\ell)].
\ee
as for transforming the Hamiltonian. The ensuing differential
equations are given in the Appendix \ref{app:flowobs}.
Eventually, we are in the position to determine
the DSF at zero temperature
and to compare it quantitatively with the measured counting
rates of inelastic neutron scattering.

\subsection{Truncation according to scaling dimension}

In general, the commutator in the flow equation \eqref{eq:flow_equ}
generates operator terms that are not present in the initial Hamiltonian. 
For interacting many-body systems the exact treatment leads generically 
to an infinite number of operator terms in the Hamiltonian and 
in the observables during the flow. This is not tractable in practice.

In order to obtain a closed set of differential equations, it is necessary to truncate in some way. For a physically justified and 
systematically controlled truncation one has to classify the relevance of the
different operator terms. A standard approach is to use a small expansion
parameter, see for instance Refs.\ \cite{knett00a,krull12}.
For gapless systems such as long-range ordered quantum magnets whose
elementary excitations are Goldstone bosons such a small parameter is
not obvious. Common choices are to expand in powers of $1/S$ or of $1/N$
(here $N$ is the number of flavors) \cite{auerb94,chubu94}. 
We tried the $1/S$ expansion, but found it inefficient.
On the one hand, it is necessary to include complicated hexatic terms,
i.e., terms with six magnon operators, in order to be consistent.
On the other hand, one does not capture the important renormalization
of the magnon-magnon interactions, see below.
Hence, we opt for the scaling dimension instead as truncation criterion
\cite{powal15}.

Originally, the concept of scaling dimension was introduced to describe critical phenomena by means of renormalization group approaches
and conformal field theories \cite{fried84,cardy96}.
It is designed to focus on the low-energy physics of a model.
 In the context of continuous
transformations, it was previously used to treat
the one-dimensional sine-Gordon model. The terms of the
operator product expansion of the vertex functions were classified
according to their scaling dimension \cite{kehre99,kehre01}.

Yet one may wonder why a concept designed to describe
the low-energy physics is appropriate for describing
magnons at short wavelengths, i.e., at high energies?
The argument runs as follows.
For gapless magnons the energy threshold 
above which the multi-particle continua start, i.e., 
the lower band edge, is given by the single magnon dispersion.
This is easy to see if the gapless mode occurs at zero momentum
in the MBZ because the lower band edge of two-magnon states at momentum
$\mathbf{q}$ is given by $\omega_2(\mathbf{q})=
\min_{\mathbf{k}}[\omega(\mathbf{q}-\mathbf{k}) + \omega(\mathbf{k})]$
which cannot be higher than $\omega(\mathbf{q})$.
In fact, it equals $\omega(\mathbf{q})$ because otherwise the
single magnon would decay into two-magnon states \cite{fisch10,fisch11,zhito13}
for which there is no evidence \cite{sandv01}.
Due to the bipartiteness of the square lattice, the single magnon
cannot decay into two magnons, but may decay in at least three magnons,
see the hybridization terms in Eq.\ \eqref{eq:hybrid}. But this modifies
the above argument only slightly such that the lower band edges
of the three-magnon continuum reads
\be
\omega_3(\mathbf{q})=
\min_{\mathbf{k}_1,\mathbf{k}_2}\left[\omega(\mathbf{q}-\mathbf{k}_1-\mathbf{k}_2) + 
\omega(\mathbf{k}_1)+ \omega(\mathbf{k}_2)\right] \quad .
\ee
Again, the continuum states close in energy to the dispersion
are those where $\mathbf{k}_1$ and $\mathbf{k}_2$ are close to zero where
the magnons are gapless. Hence, the hybridization of
one-magnon and three-magnon states is strongly influenced
by the physics at and close to zero wavevector.
Thus, the scaling dimension  is indeed an appropriate 
criterion for the relevance of certain physical processes.

If interaction and hybridization are taken into account, the multi-magnon states 
will be renormalized predominantly by the long wavelength excitations.
We have to find the terms which remain important, if one zooms to smaller 
and smaller energies or momenta, respectively. This is expressed
by the scaling dimension. We classify the relevance of operator terms by their scaling properties under the momentum transformation 
$\mathbf{k}_i \rightarrow \lambda \mathbf{k}_i$ with $\lambda<1$
in the thermodynamic limit. We consider a generic term in $D$-dimensional
momentum space
\be
\label{eq:scaling}
 \mathcal{T} \!\!= \!\!\int_{\textnormal{BZ}} 
C_{\mathbf{k}_1\ldots\mathbf{k}_{n}}
\mathcal{O}^n_{\mathbf{k}_1\ldots\mathbf{k}_{n}} 
\delta(\mathbf{k}_1+\ldots+\mathbf{k}_n)
d^D\mathbf{k}_1\ldots d^D\mathbf{k}_n ,
\ee
where $\mathcal{O}^n_{\mathbf{k}_1,\ldots\mathbf{k}_{n}}$ is a monomial of $n$ bosonic
operators of creation or annihilation type. Conservation of momentum is ensured by the $\delta$-function $\delta(\bf{k}_1+\ldots+\bf{k}_n)$. 

If the momenta are rescaled one obtains the factor 
$\lambda^{D(\frac{n}{2}-1)}$ in front of $\mathcal{T}$ 
which is the dimensional contribution of the operator term. Moreover, one has to take into account the scaling properties 
of the coefficient 
$C_{\lambda\bf{k}_1\ldots\lambda\bf{k}_{n}} \rightarrow \lambda^c C_{\bf{k}_1\ldots\bf{k}_{n}}$ at small momenta defined by its
leading behavior in the vicinity of $\bf{k}_i=0$.
Consequently, the operator term $\mathcal{T}$ acquires a total pre-factor 
$\lambda^{D(\frac{n}{2}-1)+c}$ where $D$ is the dimension, i.e.,
for the square lattice $D=2$. 
Thus, the scaling dimension $d$ of $\mathcal{T}$ is defined by the exponent 
$d=(n-2)+c$. Obviously, operator terms with a larger scaling dimension become 
less important upon scaling with $\lambda<1$.  
The crucial corollary is, that the relevance of $\mathcal{T}$ decreases 
with increasing number of its creation and annihilation operators. Hence,
interactions and hybridizations between subspaces 
with higher and higher magnon number are less and less relevant. 

For instance, the dispersion is given by
$\omega = v_\mathrm{S}\left|\bf{k}\right|$ for $\bf{k}\ll1$
where $v_\mathrm{S}$ is the spin wave velocity.
Hence, the single-magnon terms have scaling dimension $d=1$.
The quartic operator terms have scaling dimension $d=2$ because the vertex functions 
\cite{uhrig13} are bounded, i.e., $c=0$.
Hexatic terms have even higher scaling dimension 3 or higher, depending
on $c$. This is the reason why we neglect them altogether 
in the present treatment.

In this article, we do not only track the flow of the 
Hamilton operator, but also the flow of the spin observables,
which represent the vertex corrections. Thus the flow of the observables
has to be truncated as well. The corresponding
spin operators are not evaluated directly, but as a part of a resolvent
at zero temperature. 
It has turned out that it is an appropriate choice to use the number of
excited (or de-excited) magnons as criterion for the truncation. 
This is in-line with
our approach for the Hamiltonian where the scaling dimension
essentially limits the number of bosonic operators in each monomial.
For the transversal channel, in which the number of magnons is odd,
we compute the one- and three-magnon channel. In the longitudinal
channel where the number of magnons has to be even we study
the two-magnon channel. This choice is strongly supported
by the sum rules which agree very well with their
theoretical values, see below. If we had omitted an important
contribution the sum rules would indicate missing weight.

\section{Spectroscopic signatures of interacting spin waves}
\label{sec4}

Here we perform a detailed comparison between the theoretical results and the experimental data from inelastic scattering of polarized neutrons \cite{chris07a,dalla15}.
First, we consider the applicability of the simple Heisenberg Hamiltonian
\eqref{eq:hamiltonian}. It is established that in the parent compounds
of high-temperature superconductors further next-leading terms matter, see
Ref.\ \cite{majum12a} and references therein,
for instance ring exchange terms. But these terms are smaller by a factor
of $t^2/U^2$ if $t$ is the hopping element and $U$ the on-site repulsion in
an underlying Hubbard model. Since in the organic cuprate studied in 
Refs.\ \cite{chris07a,dalla15} the leading exchange $J$ is smaller
than the exchange in the high-temperature cuprates by a factor 15 to 20,
the hopping element $t$ must be smaller by about a factor 4. Hence the relative
significance of such higher terms is suppressed by a factor 15 to 20 so 
that it ranges only on the percent level. For this reason, we do not consider
it here, bearing in mind that quantitative conclusions may be influenced on 
the percent level.

\begin{figure}[htb]
\begin{center}
\includegraphics[width=0.6\columnwidth]{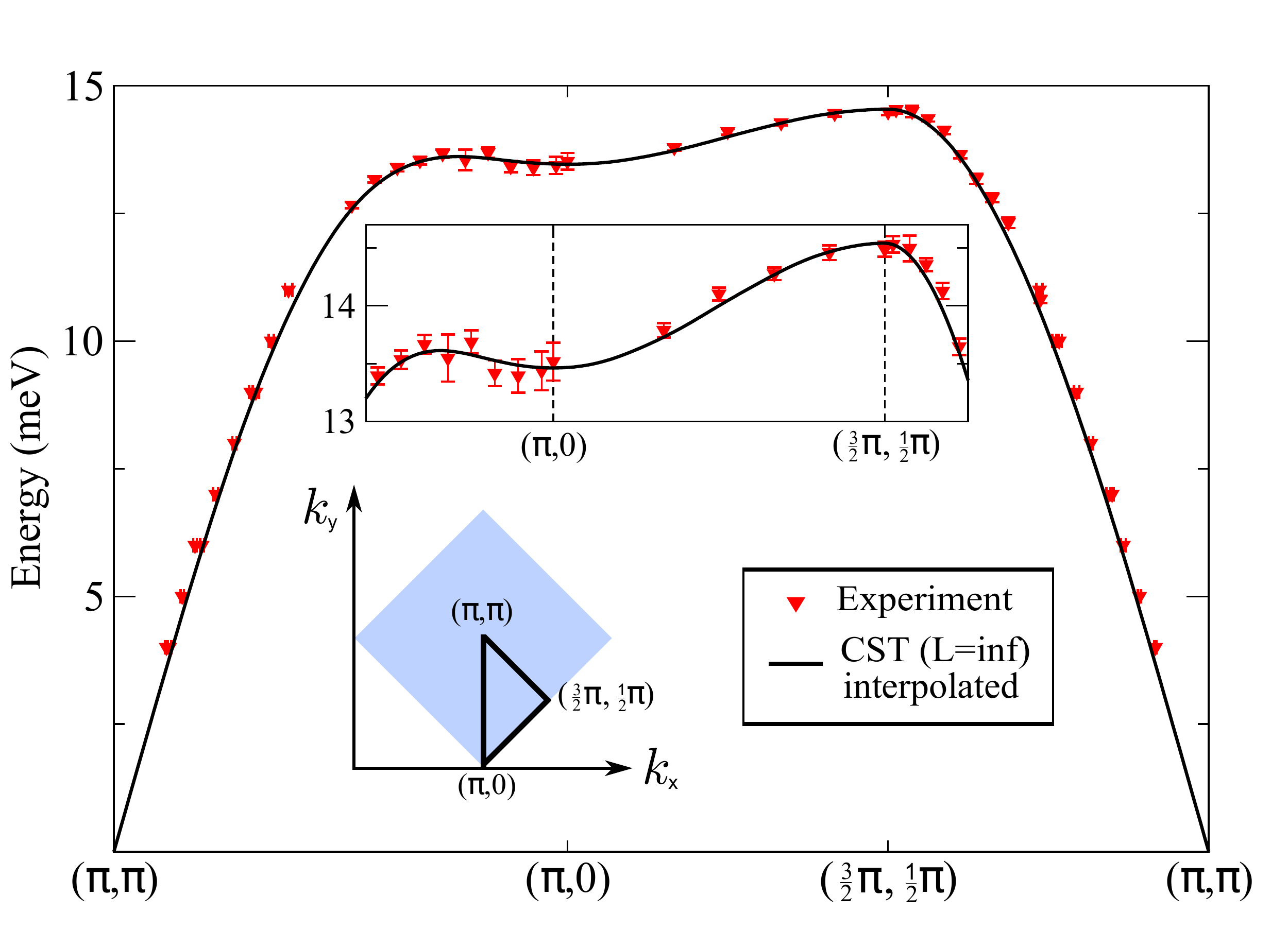}	  
\end{center}
\caption{ 
\label{fig:dispersion}
\textbf{One-magnon dispersion.} 
Dispersion along a representative path through the 
magnetic Brillouin zone (in light blue) comparing
the dispersion obtained theoretically by CST with the experimental data
(red triangles with error bars) from Ref.\ \cite{dalla15}
for the coupling value $J=6.11(2)$~meV.
The inset shows the convincing
agreement around the roton minimum at $(\pi,0)$.
The theoretical data are extrapolated to infinite system size on a mesh 
of wave vectors and interpolated between them which yields the solid line.
(The lattice constant is set to unity.) 
}
\end{figure}

Second, we determine the concrete value of the exchange coupling 
$J$, which defines the global energy scale.
This can be done by fitting the energies at 
$\mathbf{k}=(\pi,0)$  and $\mathbf{k}=(\frac{\pi}{2},\frac{\pi}{2})$ 
yielding the value $J=6.11(2)$~meV. 
The obtained excellent overall agreement of the dispersion is illustrated along 
a path along the high-symmetry lines
of the Brillouin zone in Fig.\ \ref{fig:dispersion}.

%Figure: Dispersion
%%%%%%%%%%%%%%%%%%%

\begin{figure}[htb]
\begin{center}
\includegraphics[width=0.6\columnwidth]{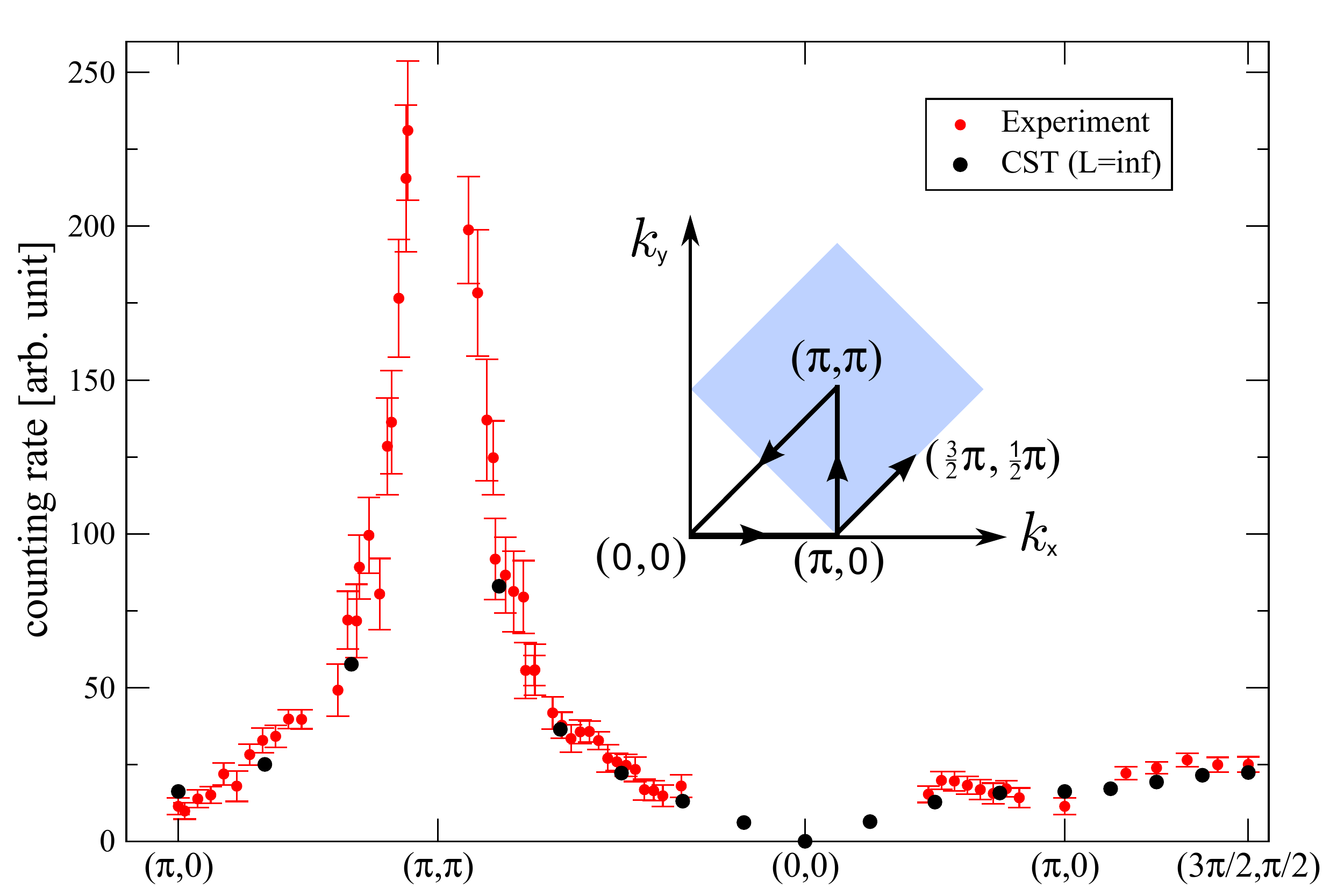}  
\end{center}
\caption{ 
\label{fig:weights}
\textbf{Transversal spectral weight of the single magnon excitations.} 
Comparison of the measured weights from Ref.\ \cite{chris07a}
with the computed weights $S_1^{xx}(\mathbf{Q})+S_1^{yy}(\mathbf{Q})$, 
i.e., the transversal SSF}, see Eq.\ \eqref{eq:static}. 
An overall proportionality factor is fitted.
\end{figure}

We argued above that it is crucial to transform the observables as well.
This means that the vertex corrections need to be taken into account.
Fig.\ \ref{fig:weights} compares the data obtained in this way with
experimental results in the transversal channel. The agreement is very good 
and corroborates that the renormalized spin wave theory including the
transformation of observables captures the relevant physics.
This is supported further by the sum rules \cite{zheng05}.
We checked the sum rules for $S=1/2$ in the transversal channel
\be
S_\textscript{t}^{\textscript{tot}} =
\frac{1}{2}\int_{\textnormal{BZ}} d\mathbf{Q}  \left(S^{xx}(\mathbf{Q})
+ S^{yy}(\mathbf{Q})\right) 
= S \approx 0.495
\ee
and in the longitudinal channel
\be
S_\textscript{l}^{\textscript{tot}} = 
\int_{BZ} d\mathbf{Q} S^{zz}_{\text{eff}}(\mathbf{Q}) 
= S^2 \approx  0.2403.
\ee
Note that the above integrals run over the full Brillouin zone, not the MBZ.
The numerical values are obtained by extrapolating the effective observables
 to the thermodynamic limit. We emphasize that the nice agreement indicates
that it is indeed quantitatively sufficient to consider the one-magnon
and three-magnon contributions in the transversal channel and the two-magnon
contribution in the longitudinal channel.

%Figure: Weights
%%%%%%%%%%%%%%%%%%%
%%%%%%%%%%%%%%%%%%%

\begin{figure*}
\begin{center}
\includegraphics[scale=0.30]{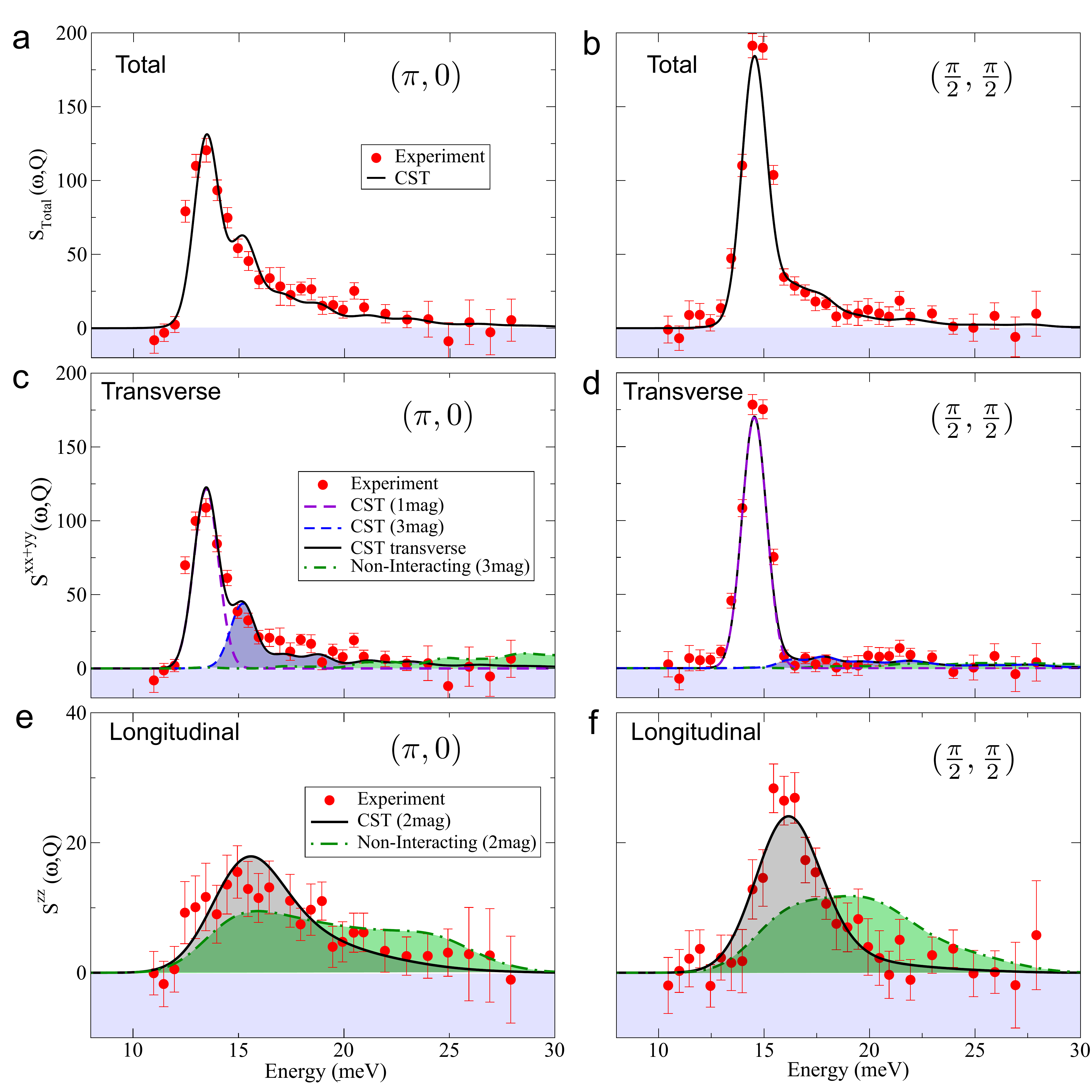}  
\end{center}
\caption{ 
\label{fig:shapes} 
\textbf{Dynamic structure factors.} 
Comparison between the DSFs measured in Ref.\ \cite{dalla15}
and the theoretical line shapes obtained from the CST. 
The energy scale is set to $J=6.11(2)$meV for all curves.
The arbitrary units on the $y$-axis are fixed globally 
to match the experimental data. The lattice constant is set to unity. 
(\textbf{a})~Total DSF (sum of transversal and longitudinal part) at $\k =(\pi,0)$.
(\textbf{b})~Total DSF at $\k =(\frac{\pi}{2},\frac{\pi}{2})$.
(\textbf{c})~Transversal DSF at $\k =(\pi,0)$.
The magenta line shows the dominant one-magnon peaks.
The transversal broadening 
is set to \mbox{$\sigma_\text{t}=0.58(2)$meV}.
The three-magnon continuum is shown as blue curve. Omitting
the interaction leads to the green curve which does not agree
with experiment at all. (\textbf{d})~Transverse DSF at
$\k =(\frac{\pi}{2},\frac{\pi}{2})$; otherwise same as in panel (\textbf{c}).
(\textbf{e})~Longitudinal DSF at $\k =(\pi,0)$. 
The longitudinal broadening is set to
$\sigma_\text{l}=1.41(5)$meV to match the experimental data.
The black line depicts the two-magnon continuum; the four-magnon
contribution is negligible. The green line results from
omitting the magnon-magnon interaction. (\textbf{f})~Longitudinal
DSF at $\k =(\frac{\pi}{2},\frac{\pi}{2})$; otherwise same as 
in panel (\textbf{e}).}
\end{figure*}

The main results for the DSFs are depicted in Fig.\ \ref{fig:shapes} 
where the dynamics in multi-magnon channels enters.
To obtain these results we proceeded as follows, see also the Appendix 
\ref{app:B}.
The experimentally measured counting rates in the transversal channel 
at $\mathbf{k}=(\pi,0)$ and $\mathbf{k}=(\frac{\pi}{2},\frac{\pi}{2})$ 
(panels (c) and (d) in Fig.\ \ref{fig:shapes}) exhibit pronounced peaks which we identify as the one-magnon peaks at the energies given by the dispersion 
$\omega_{\mathbf{k}}^{\text{eff}}$ in \eqref{eq:eff_single_spinwaves}.

Next, we address the broadening $\sigma$ on which the line shapes depend.
We choose Gaussian broadening and convolve the theoretical results with 
the Gaussians to mimic various broadening mechanisms such as 
the finite experimental resolution, the uncertainties in the background subtraction,
and finite residual temperature and disorder effects, see Eq.\
\eqref{eq:gaussian_sd}. 
Thus, the transversal broadening $\sigma_\text{t}$ and the longitudinal broadening
$\sigma_\text{l}$ enter as additional fit parameters. 
The broadening and overall height of the line shapes is fitted to match 
the measured counting rates.
The broadening is set to $\sigma_\textscript{t}=0.58(2)$~meV in the transversal channel and 
to $\sigma_{\textscript{l}}=1.41(5)$~meV in the longitudinal channel. We recall,
that the units of the energy axis and the counting rates are fixed for all 
displayed panels for both, the transversal and longitudinal channel.

%%%%%%%%%%%%%%%%%%%
%%%%%%%%%%%%%%%%%%%

The very good overall agreement between the theoretical curves and 
the experimental data in Figure~\ref{fig:shapes} is convincing.
First, we address the total DSF being the sum of the transversal and 
longitudinal DSFs. The positions and the heights of the 
pronounced one-magnon peaks as well as the continuum tails are captured by the theoretical line shapes in a quantitative way. The slight wiggles in the 
continua are due to the finite discretization in the Brillouin zone.

In the transversal channel one can discriminate the one-magnon contribution
given by a Gaussian function (dashed magenta line) and the three-magnon continuum shaded in blue. An important experimental feature is the pronounced continuum 
tail at $\mathbf{k}=(\pi,0)$. By contrast, the continuum a 
$\mathbf{k}=(\frac{\pi}{2},\frac{\pi}{2})$ is much weaker; its spectral weight is 
marginal. Both aspects are captured with remarkable accuracy by the 
theoretical line shapes. The spectral weight in absolute units
in the one-magnon channel is extrapolated for $L=\infty$ to be $0.5839$ at 
$\mathbf{k}=(\frac{\pi}{2},\frac{\pi}{2})$  and $0.4339$ at $\mathbf{k}=(\pi,0)$.
The spectral weight in the three-magnon
channel is extrapolated for $L=\infty$ to be $0.1337$ and $0.2952$, respectively.
This implies that at $\mathbf{k}=(\frac{\pi}{2},\frac{\pi}{2})$
$81.4\%$ of the weight rest in the one-magnon peak while at 
$\mathbf{k}=(\pi,0)$ it is only $59.5\%$ due to the hybridization with
the three-magnon continuum.
Previously, the significant continuum at $\mathbf{k}=(\pi,0)$
 was interpreted as an indication of a 
fractionalization into  spinons~\cite{dalla15}.
Further results on spectral weights in the various channels are
given in the Appendix \ref{app:specweights}.

To illustrate the crucial relevance of the magnon-magnon interaction, we determine 
the three-magnon continuum in the non-interacting case as well, i.e., we omit
the interaction $\mathcal{V}^0$ in the effective model
while leaving everything else unchanged.
This yields the green curves.
The difference between the magnon continuum for the interacting and 
non-interacting case at $\mathbf{k}=(\pi,0)$ is striking.
If the magnon-magnon interaction is omitted the spectral weight is shifted to 
significantly higher energies leading to a clear mismatch with the experimental findings. 

The observations in the longitudinal channel are similar. The spectral weight
in the two-magnon channel is found to be $0.2508$ at 
$\k =(\frac{\pi}{2},\frac{\pi}{2})$
and $0.2457$ at $\k =(\pi,0)$. The CST results agree 
very well with the experimental data. The very good accord is obviously spoiled 
by omitting the magnon-magnon interaction. Consequently, the pronounced signals in
the measured intensities can be directly identified as the longitudinal magnon 
or Higgs resonance. The resonance is still fairly broad which implies
that the longitudinal magnon lives only for a short time. The short life
time can be traced back to the fact that the energy of the longitudinal magnon
lies right within the two-magnon continuum into which it decays.

The important spectral weight found in the high-energy tails of the
transversal spectral response results from the three-magnon continuum.
This is a pivotal point because previous analyses take
the continua as smoking gun of the relevance of a spinon scenario \cite{ho01,dalla15}.
Continuous contributions from three magnons only arise in the DSF from the vertex corrections,
i.e., from the transformation of the spin observables. The hybridization
terms have to be involved at least once in the CST of the observables.
Otherwise, the Dyson-Maleev representation does not allow for a physical process 
linking two scattering states comprising three magnons each.
This underlines the crucial progress achieved in the present work
by the CST of the observables in comparison
with the previous analysis \cite{powal15}.

The distribution of spectral weight \emph{within} each continuum 
is a direct consequence of the attractive magnon-magnon interaction.
The attractive magnon-magnon interaction shifts spectral weight \emph{within} the 
three-magnon continuum to lower energies. By means of level repulsion,
this in return decreases  the energy of the single magnon states, i.e., the
dispersion. This is the physical origin of the so-called roton dip in the 
dispersion at $\k =(\pi,0)$. Qualitatively, this physics is also
be found by describing the Higgs resonance by 
so-called singlons \cite{akter16a}. We see that the size of the 
renormalized magnon-magnon interaction is very important. Thus,
its quantitative renormalization matters and we pointed out 
previously, that it is enhanced by about 50\% in the renormalization
flow, see Supplement of Ref.\ \cite{powal15}.

To back the above given physical scenario we present a quantitative analysis 
of the interplay between the magnon-magnon attraction and the 
hybridization effects in the next section.

%%%%%%%%%%%%%%%%%%%%%%%
%%%%%%%%%%%%%%%%%%%%%%%
\section{Non-perturbative renormalization of high-energy magnons}
\label{sec5}
%%%%%%%%%%%%%%%%%%%%%%%%%%%%%%%%%%%%%%%%%%%%%%%%

Each type of operator term stands for a particular physical process.
The solution of the flow equations, given in the Appendix \ref{app:A},
tells us how the prefactor of such a 
physical process and hence its significance changes in the course of the 
flow, i.e., how it is renormalized. We use this information to determine
the relevance of different physical processes and how they influence
the effective Hamiltonian \eqref{eq:eff_sw_hamiltonian} 
at the end of the continuous unitary transformation.
An important example is the hybridization between one- and three-magnon states.
We find that the hybridization between one and three-magnon states contributes
significantly to the renormalization of the anomalous high-energy dispersion.  

\begin{figure}[htb]
\begin{center}
\includegraphics[width=0.6\columnwidth]{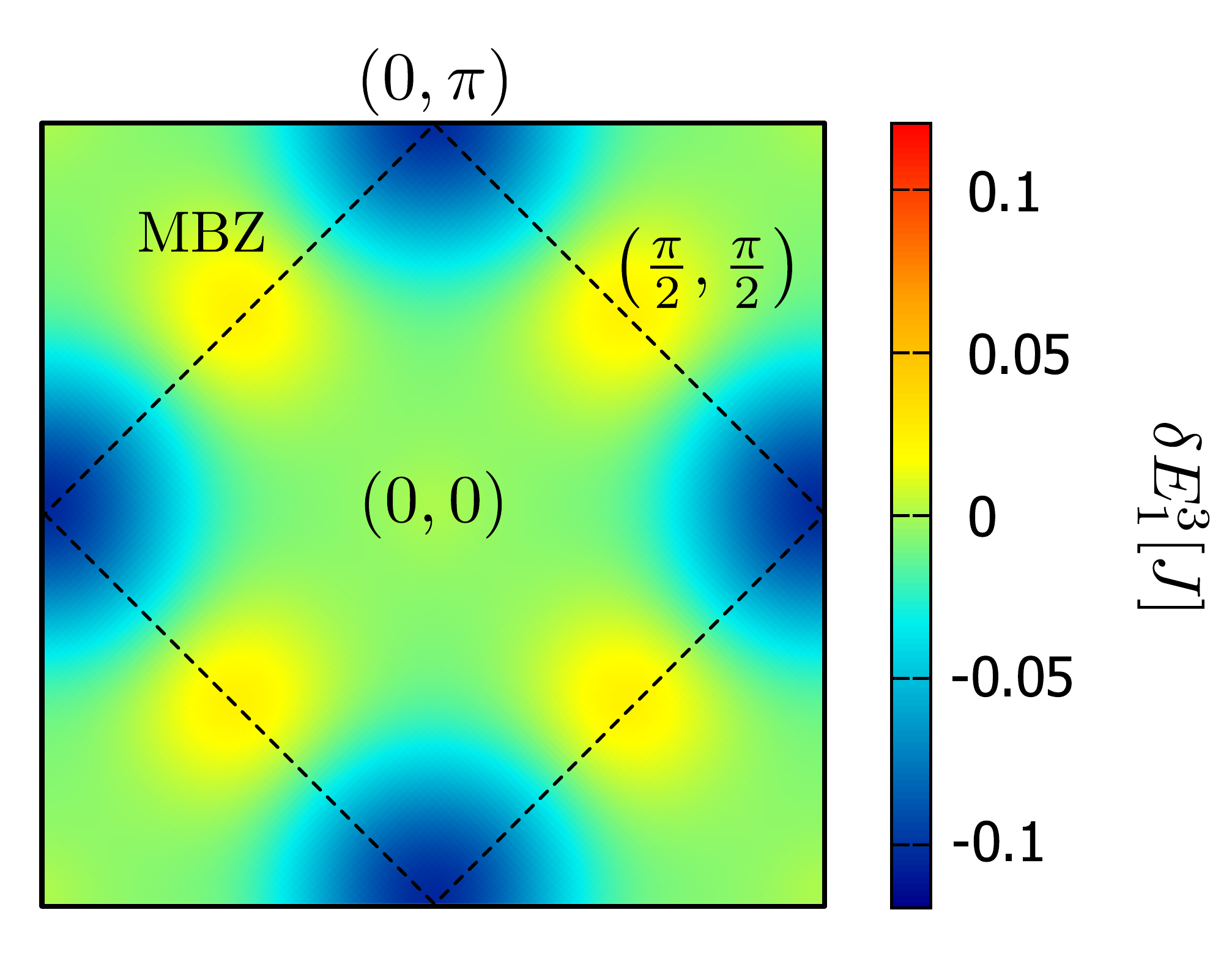}  
\end{center}
\caption{Shift of the dispersion given by \eqref{eq:hybridization} due to the 
flowing, i.e., renormalized hybridization terms $\mathcal{V}^\pm$. Note that the
shift is positive at wavevectors $\k=(\pm\frac{\pi}{2},\pm\frac{\pi}{2})$
while it is clearly negative at $\k=(\pm\pi,0)$ and $\k=(0,\pm\pi)$.
\label{fig:disp-shift} 
} 
\end{figure}

The contribution of the hybridization to the 
one-magnon dispersion is given by the term
\be
\label{eq:hybridization}
  \delta E_1^3(\k) = -4\int_{0}^{\infty} d\ell
	\sum_{\mathbf{1}, \mathbf{2}, \mathbf{3}} 
	\delta^{\mathbf{1}\mathbf{2}}_{\k-\mathbf{3}} 
	\mathcal{V}_{\k\mathbf{\uminus3}\mathbf{2}\mathbf{1}}^{(\alpha \uminus 3)}(\ell)
	\mathcal{V}_{\mathbf{1}\mathbf{2}\mathbf{\uminus3}\mathbf{k}}^{(3\uminus\alpha )}
	(\ell)
\ee
 which stems from the corresponding summand in the flow of the one-magnon dispersion 
$\partial_{\ell}\omega_\k(\ell)$. The resulting shift of the
dispersion is depicted in Fig.\ \ref{fig:disp-shift}. Clearly, this
explains the occurrence of the roton dips because the dispersion is lifted
upward at wavevectors $\k=(\pm\frac{\pi}{2},\pm\frac{\pi}{2})$.
In contrast, it is pushed downwards at wavevectors
$\k=(\pm\pi,0)$ and $\k=(0,\pm\pi)$.

\begin{figure}[htb]
\begin{center}
\includegraphics[width=0.9\columnwidth]{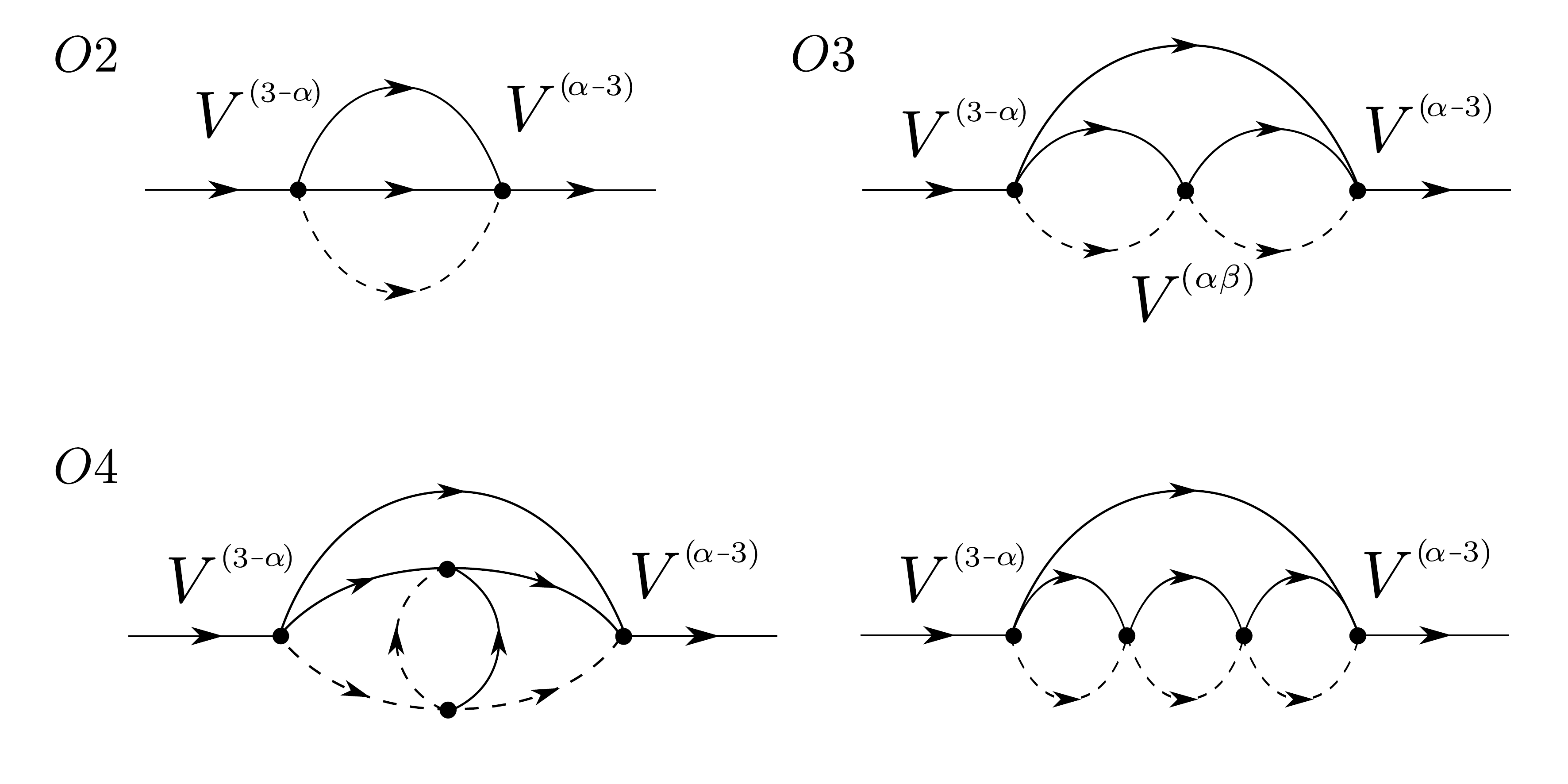}  
\end{center}
\caption{Three contributions $\delta E_1^3(\k)$ from the hybridization terms
in \eqref{eq:hybrid} to the renormalized one-magnon dispersion 
$\omega_\k^{\text{eff}}$  as given in \eqref{eq:hybridization}.
The solid lines stand for the propagation of an $\alpha$-magnon while
the dashed lines stand for the propagation of a $\beta$-magnon.
The upper diagram is second order ($O2$) in the quartic terms $V$. 
The second diagram is third order ($O3$) in the quartic terms $V$;
the additional order results from an interaction between an $\alpha$ 
and a $\beta$ magnon.  The third diagram is fourth order ($O4$) in the quartic terms $V$; the two additional order result from two hybridization vertices \eqref{eq:hybrid}. The lower right diagram is also of fourth order;
it represents an iterated interaction between the $\alpha$- and 
$\beta$-magnon. \label{fig:higgs-a} 
}
\end{figure}

To elucidate the significance and the origin
of $\delta E_1^3(\k)$ further we consider a 
perturbative evaluation of \eqref{eq:hybridization}
using the conventional spin wave expansion in $1/S$ first. 
Subsequently we interpret the full non-perturbative evaluation of 
\eqref{eq:hybridization}.
In leading order in $1/S$ the flow of the coefficients is given by
\bes
\label{eq:perturbative_flow}
\bea
\mathcal{V}_{\mathbf{k}\mathbf{\uminus3}\mathbf{2}\mathbf{1}}^{(\alpha \uminus 3)}
(\ell) &=&
V_{\mathbf{k}\mathbf{\uminus3}\mathbf{2}\mathbf{1}}^{(\alpha \uminus 3)}
\exp\left[-(\omega^{(0)}_{\mathbf{k}}-\omega^{(0)}_{\mathbf{1}}-
\omega^{(0)}_{\mathbf{2}}-\omega^{(0)}_{\mathbf{3}})\ell\right] 
\\
\mathcal{V}_{\mathbf{1}\mathbf{2}\mathbf{\uminus3}\mathbf{k}}^{(3\uminus\alpha )}
	(\ell) &=&
 V_{\mathbf{1}\mathbf{2}\mathbf{\uminus3}\mathbf{k}}^{(3\uminus\alpha )}
\exp\left[-(\omega^{(0)}_{\mathbf{k}}-\omega^{(0)}_{\mathbf{1}}-\omega^{(0)}
_{\mathbf{2}}-\omega^{(0)}_{\mathbf{3}})\ell\right]
\eea
\ees
so that the leading term of $\delta E_1^3(\k)$ is second order in 
$V$. Using \eqref{eq:perturbative_flow} we obtain 
\be
  \delta E_1^3(\mathbf{k})=   
	\sum_{\mathbf{1}, \mathbf{2}, \mathbf{3}}   
	\frac{2 V_{\mathbf{k}\mathbf{\uminus3}\mathbf{2}\mathbf{1}}^{(\alpha \uminus 3)}
	V_{\mathbf{1}\mathbf{2}\mathbf{\uminus3}\mathbf{k}}^{(3\uminus\alpha )} 
	\delta^{\mathbf{1}+\mathbf{2}}_{\mathbf{k}-\mathbf{3}}}
	{\omega^{(0)}_{\mathbf{k}}-\omega^{(0)}_{\mathbf{1}}-
	\omega^{(0)}_{\mathbf{2}}-\omega^{(0)}_{\mathbf{3}}} 
\ee
which equals the result of second order perturbation theory in $V$ 
specified by the upper left diagram $O2$  shown in Fig.\ \ref{fig:higgs-a}.

In third order, the upper right diagram $O3$ in Fig.\ \ref{fig:higgs-a},
the attractive interaction between an $\alpha$-magnon and a $\beta$-magnon
is included. Such a diagram is taken into account in a full perturbative 
calculation up to $1/S^3$ as carried out by Syromyatnikov \cite{syrom10}.
Further contributions lead to higher corrections such as depicted
in the diagrams $O4$ in the second row of Fig.\ \ref{fig:higgs-a}.
Interestingly, the left $O4$ diagram results from the renormalization
of the magnon-magnon interaction while the right $O4$ diagram results
from the renormalization of the hybridization terms. 
Physically, it represents the repeated 
interaction processes between an $\alpha$-magnon and a $\beta$-magnon.

\begin{figure}[htb]
\begin{center}
\includegraphics[width=0.6\columnwidth]{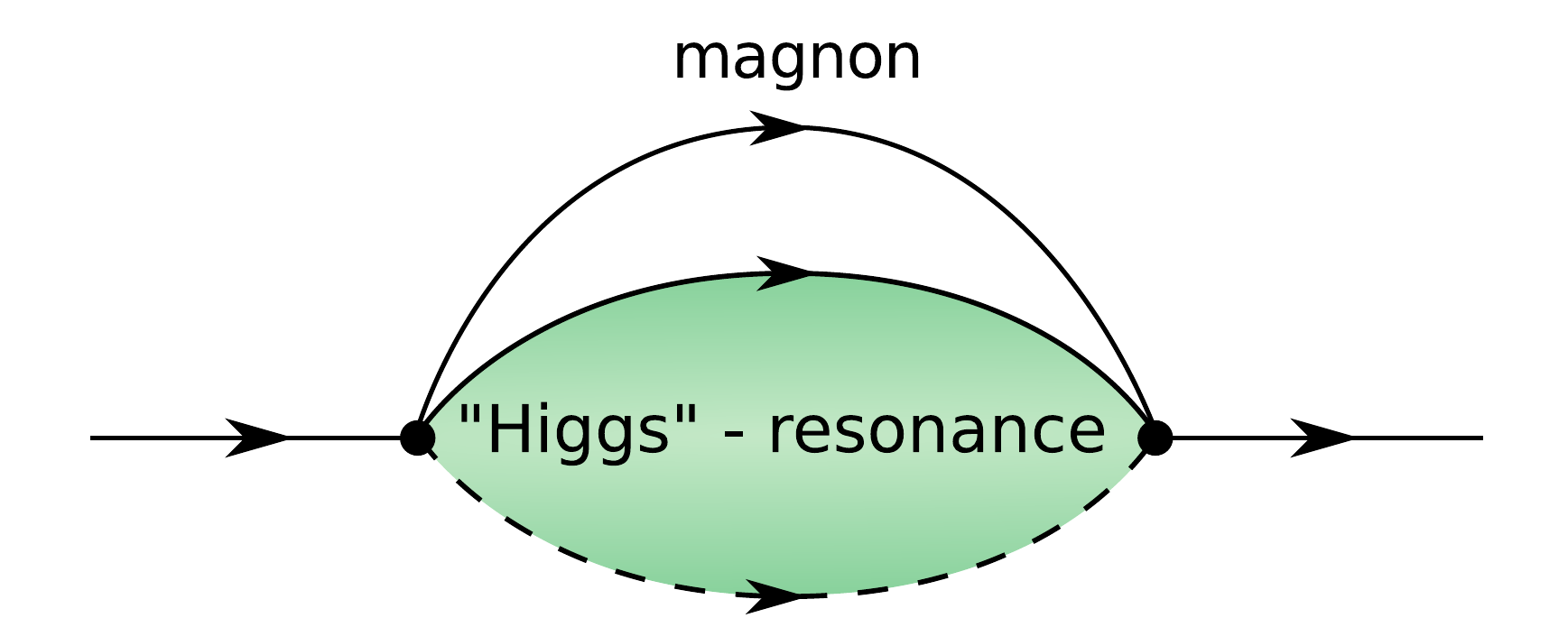}  
\end{center}
\caption{Sum of all contributions to $\delta E_1^3(\k)$ from the
renormalization of the propagation of a pair of $\alpha$- and
$\beta$-magnon which eventually form the Higgs resonance. 
\label{fig:higgs-b} 
} 
\end{figure}

Our results, in combination with
the slow convergence found in the direct perturbative approach \cite{syrom10},
indicate that the $O4$ and higher terms are quantitatively significant.
In our non-perturbative renormalizing approach the flow of all coefficients 
up to scaling dimension $d=2$ is evaluated including all
mutual dependencies. As a result, the term in 
\eqref{eq:hybridization} includes vertex corrections up to infinite order in $1/S$ as illustrated in Fig.\ \ref{fig:higgs-b}. The interaction and hybridization processes renormalize the propagation of a pair of $\alpha$- and
$\beta$-magnon which form the Higgs resonance \cite{powal15}.
We stress, however, that there are also processes linking the upper  
$\alpha$-magnon and the $\beta$-magnon and processes linking the two 
$\alpha$-magnons (both not included in the diagram in Fig.\ \ref{fig:higgs-b}). 
The latter are of minor importance, see below.  But all these processes are included 
in the solution of the flow equation \eqref{eq:flow_equ}. 

The significance of the renormalization of the attractive interaction
between an $\alpha$- and a $\beta$-magnon is strongly corroborated by the fact
that the downshift of the dispersion at the roton minimum diminishes considerably 
(from $8\%$ to $5\%$) if we switch off the flow of the interaction by 
artificially setting $\partial_{\ell} \mathcal{V}^{\alpha\beta}(\ell)=0$. 
By contrast, the renormalization of the interaction between magnons of the same sublattice (given by $\mathcal{V}^{\alpha\alpha}$ or $ \mathcal{V}^{\beta\beta}$) 
does not affect the high-energy dispersion significantly.

Finally, in order to quantify the energy reduction induced by the magnon-magnon 
interaction $\mathcal{V}^{\alpha\beta}$ we diagonalize the two-magnon channel
with one $\alpha$-magnon and one $\beta$-magnon for finite system sizes. The
 resulting spectrum is compared with the non-interacting case 
$\mathcal{H}^{\text{eff}}_{\text{SW}}(\mathbf{k})$ where the energy spectrum is
 simply given by the sum of two one-magnon energies, i.e., two values of 
the dispersion, with given total momentum $\mathbf{k}$. 
Then, we define the difference between the next-lowest energy levels 
in these two cases given by 
\be
\label{eq:diff-energy}
\Delta E^{\textscript{2-mag}}(\mathbf{k}) := 
E_1\left(\mathcal{H}^{\text{eff}}_{\text{SW}}(\mathbf{k})\right) - 
E_1\left(\mathcal{H}^{\text{eff}}(\mathbf{k})\right)
\ee 
where $E_i(\mathcal{H})$ denotes the $i$-th eigenvalue of $\mathcal{H}$ 
sorted in ascending order counting degenerate eigenenergies only once.
Note that the lowest energy level $E_0(\textbf{k})$ is defined by the one-magnon energy 
$\omega_{\mathbf{k}}^{\textscript{eff}}$ representing the lower band edge
of the two-magnon continuum. This band edge is by construction unaffected
by the interaction. Hence it does not provide any information on 
attractive forces or binding so that we resorted to $E_1$ to define 
$\Delta E^{\textscript{2-mag}}(\mathbf{k})$.

\begin{figure}[htb]
\begin{center}
\includegraphics[width=0.6\columnwidth]{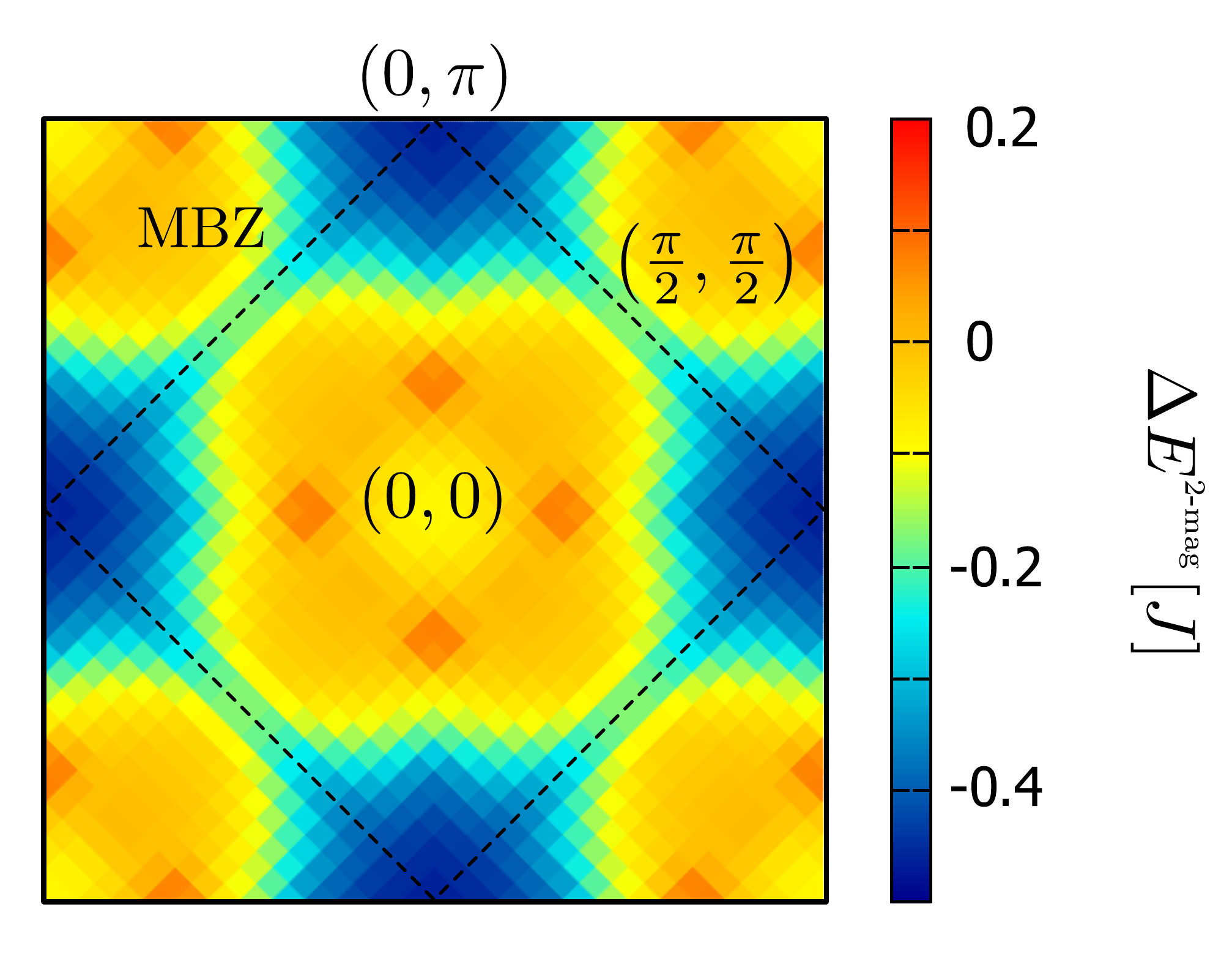}  
\end{center}
\caption{\label{fig:5bb} 
Energy difference as defined in Eq.\ \eqref{eq:diff-energy} between
the lowest energy level in the longitudinal channel, i.e., 
between one $\alpha$-magnon and one $\beta$-magnon, with and without
interaction.} 
\end{figure}

Fig.\ \ref{fig:5bb} indicates where spectral weight is
shifted downwards in the Brillouin zone and to what extent. 
Keeping the scale in mind one realizes that there is a general trend
of downshifting. But the downshifts are clearly maximum
around $\k=(\pm\pi,0)$ and $\k=(0,\pm\pi)$ and smaller around
at $\k=(\pm\frac{\pi}{2},\pm\frac{\pi}{2})$. Thus, a comparison
of Figs.\ \ref{fig:disp-shift} and \ref{fig:5bb} shows
very similar qualitative features. This observation underlines
our interpretation that one-magnon states hybridize with three-magnon states 
which are essentially built from single magnons and a Higgs resonance (or 
longitudinal magnon). As the energy of the Higgs resonance is lowered at 
$\mathbf{k}=(\pi,0)$ and its equivalent wavevectors
 the corresponding magnon-Higgs continuum is repelling the energy level of
the  single-magnon state, see Fig.\ \ref{fig:higgs-b}. 
Our results indicate that it is mandatory to
track the full renormalization of the interaction and the hybridization
terms in order to capture the physics quantitatively.

%%%%%%%%%%%%%%%%%%%%%%%
%%%%%%%%%%%%%%%%%%%%%%%
\section{Conclusions}
\label{sec6}
%%%%%%%%%%%%%%%%%%%%%%%
%%%%%%%%%%%%%%%%%%%%%%%

Summarizing, our detailed analysis shows that 
the effective magnon model obtained by renormalizing via a continuous
basis change captures the physics of the long-range ordered
Heisenberg model on a square lattice quantitatively.
This is in stark contrast to
a perturbative treatment expanding in $1/S$ which converges slowly.
Including also the appropriate renormalization of the observables, i.e.,
including the relevant vertex corrections within the CST formalism, 
allows us to obtain an impressive agreement with experiment.
This holds for both channels, transversal and longitudinal with
respect to the staggered magnetization which is the order parameter.
 
In particular, the much debated continua can be reproduced.
In order to retrieve the weight in the three-magnon continuum occurring
in the transversal DSF, the proper treatment of the transformations
of the observables is decisive. For the distribution of the weights
in the two-magnon longitudinal and in the three-magnon transversal 
continua it is important to take the renormalized attractive 
magnon-magnon interaction properly into account.
To capture the continua properly is a crucial aspect because previously the 
significant continua were interpreted as evidence
for a failure of the magnon description, justifying
to search beyond the Goldstone bosons for 
qualitatively different fractional excitations such as spinons.
We stress that our approach yields a very good agreement with
experimental data without resorting to spinons.

In conclusion, the Goldstone bosons of the long-range
ordered two-dimensional Heisenberg model not only describe
the dynamics at low energies, but also at high energies
if the magnon-magnon interaction is taken into account quantitatively. 
Nevertheless, the magnon picture of the square-lattice Heisenberg model seems to be fragile to additional four-spin interactions as 
recent quantum Monte Carlo simulations suggest \cite{shao17}. 
Thus, the authors regard the high-energy magnons close to the roton 
minimum as pairs of nearly deconfined spinons in the pure square-lattice
Heisenberg model. We stress that this characterization is not 
in contradiction to our findings. It would be very interesting to study the influence of such four-spin interactions on the magnons with our 
CST formalism.
 
With respect to high $T_c$ superconductivity, the pure magnetic side
of the long-standing quest for understanding
the underlying physics appears to be solved.
This provides a firm basis to tackle the ensuing hole-magnon 
interaction and the induced hole-hole interaction 
in future studies.

In view of the wide-spread presence of 
long-range magnetic order in general,
the continuous similarity transformation introduced here in great detail
provides a powerful tool to study the dynamics of the elementary
excitations, the Goldstone magnons, in such systems. 
This includes the important effective magnon-magnon interaction as well
as the vertex corrections describing the effective observables. Consequently, 
we expect future applications to a variety of fascinating physical systems 
such as two- and three-dimensional quantum magnets close to quantum criticality where 
the Higgs amplitude plays an even more important role in the dynamical correlation functions 
\cite{jain17,hong17,lohof15,lohof17,qin17}.

\FloatBarrier

%%%%%%%%%%%%%%%%%%%%%%%%%%%%%%%%%%%%%%%%%%%%%%%%%%%%%%%%%%%%%%%%%%%%%%%%%%%%%%%%
% Acknowledgement
%%%%%%%%%%%%%%%%%%%%%%%%%%%%%%%%%%%%%%%%%%%%%%%%%%%%%%%%%%%%%%%%%%%%%%%%%%%%%%%%
\section*{Acknowledgements}
This work was supported by the Cusanuswerk (MP) and by the Deutsche
Forschungsgemeinschaft and the Russian Foundation of Basic Research
in TRR 160. We thank N.\ Christensen, B.\ Normand, 
H.\ R\o{}nnow, A.\ Sandvik, R.\ Singh, and A.\
Syromyatnikov for fruitful discussions and exchange of data.

%\bibliography{../../../bibinput/liter10}

\begin{appendix}

\section{Derivation of the effective magnon description}
\label{app:A}

In the following part, we explicate technical details 
concerning the derivation of the effective magnon description for the Heisenberg
quantum antiferromagnet with $S=1/2$ on the square lattice using a particle 
conserving continuous  similarity transformation (CST).

\subsection{Dyson-Maleev representation}

The transformation itself has been given in the main text in Eq.\ 
\eqref{eq:dyson_maleev_representation} and
the resulting Hamilton operator in Eqs.\ 
\eqref{eq:initial_sw_hamiltonian} to \eqref{eq:hybrid}.

\subsubsection{Observables}

\begin{figure}[ht]
	\centering
		\includegraphics[width=0.4\textwidth]{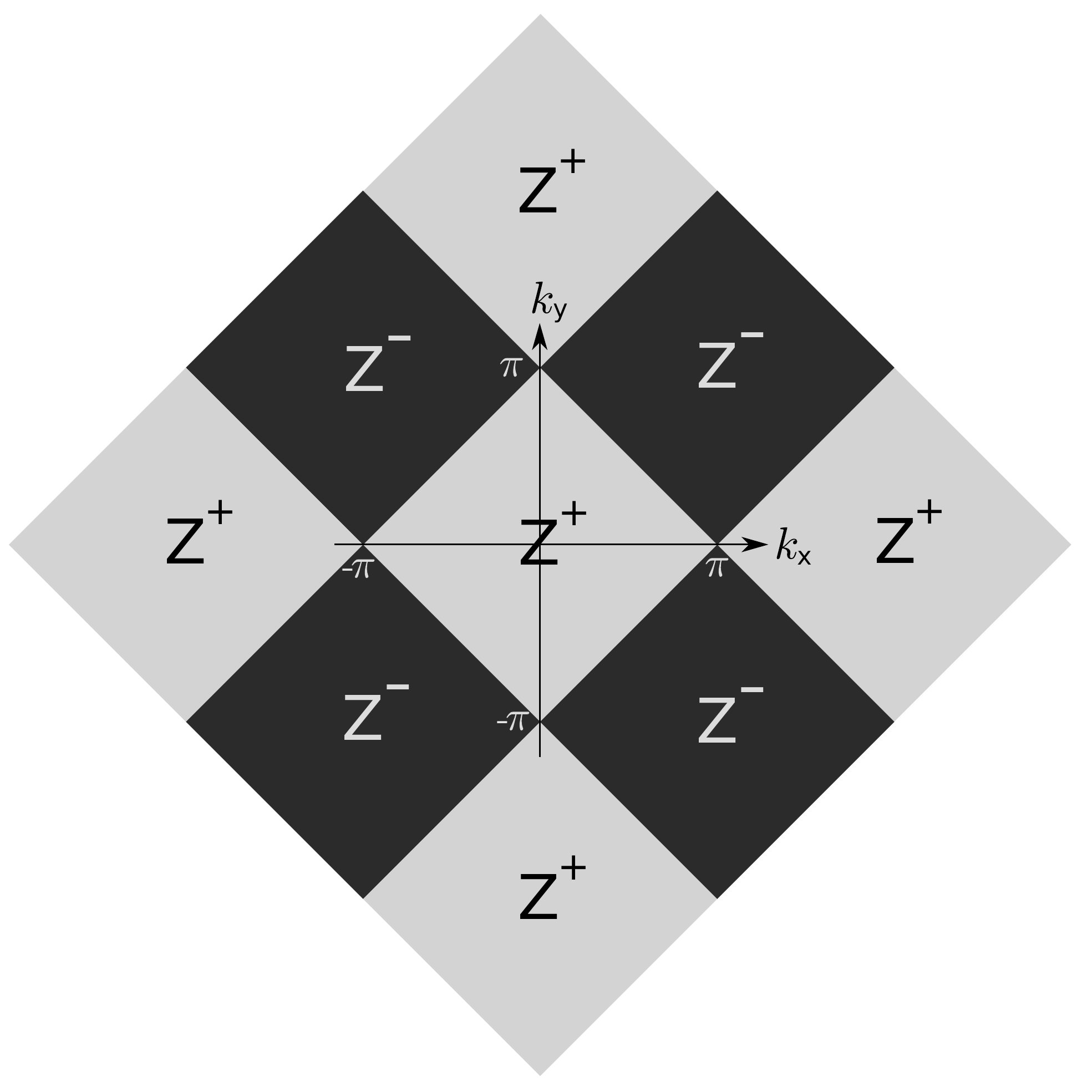}
	\caption{Illustration of the sign structure of the factor $\Gamma_{\bf{K}}$. The 
	factor takes the value $\Gamma_{\bf{K}}=+1$ for $\bf{K}\in Z^+$ or 
	$\Gamma_{\bf{K}}=+1$ for $\bf{K}\in Z^-$.}
	\label{fig:sign_MBZ}
\end{figure}

The operators which are required for the evaluation of the dynamic structure factors are obtained by the transformation in Eq.\ \eqref{eq:dyson_maleev_representation} of the main text.
For the longitudinal part, we consider the operator
\begin{eqnarray}
	      	S^z(\bf{Q}) &=& S_{A}^z(\bf{Q}) + S_{B}^z(\bf{Q}) = 
						\left( m_{Q}^2 -S N\right)\left(\Gamma_{\bf{Q}}-1 \right)\dG{\bf{Q}} 
						\nnp						
						 &+& \sum_{\bf{1},\bf{2}}  \dG{\bf{Q} - \bf{1} +\bf{2}} \Bigl\{
						 \left[\Gamma_{\bf{K}} l_{\bf{1}} l_{\bf{2}} 
						- m_{\bf{1}} m_{\bf{2}}\right] \bed{1}\beo{2}  \Bigr. 	
						+\, \left[\Gamma_{\bf{K}} m_{\bf{1}} m_{\bf{2}} - 
						l_{\bf{1}} l_{\bf{2}}\right] \ald{1}\alo{2} \nn
								&+& \left[\Gamma_{\bf{K}} m_{\bf{1}} l_{\bf{2}} 
								- l_{\bf{1}} m_{\bf{2}} \right] \ald{1}\bed{\uminus 2}
							+\, \left[\Gamma_{\bf{K}} m_{\bf{2}} l_{\bf{1}} - l_{\bf{2}} m_{\bf{1}} \right] \alo{2} \bo{\uminus 1}\Bigr\}	
\end{eqnarray}
with the factor $\Gamma_{\bf{K}} = \gamma(\bf{Q} - \bf{1} +\bf{2})$ which tracks the
 reciprocal lattice vector $\bf{K}=\bf{Q} - \bf{1} +\bf{2}$. Depending on the MBZ in
 which this vector $\bf{K}\in \Gamma_A^*$ is located the function takes the values 
$\Gamma_{\bf{K}} =1$ or $\Gamma_{\bf{K}} =-1$. It is positive in the first MBZ and negative in the adjacent edge-sharing MBZs, i.e., it switches sign each time one 
enters another MBZ across an edge as shown in Fig.~\ref{fig:sign_MBZ}.

For the transversal part we transform the operators
\begin{eqnarray}
  S^{+}(\bf{Q}) &=& 
		\sqrt{2SN} \left( 1 -  \frac{1}{S N} \sum_{\bf{1}} m_{\bf{1}}^2 \right) 
	   \left( \left(l_{\bf{Q}} + {\Gamma_{\bf{ Q}}} \, m_{\bf{Q}}\right)\, \alo{-Q} 
		+ \left(m_{\bf{Q}} + {\Gamma_{\bf{ Q}}} \, l_{\bf{Q}}\right) \, \bed{\bf{Q}} 
		\right) \nn
 &{-}& \frac{1}{\sqrt{2SN}}  \sum_{\bf{1,2,3}} \, \dG{\bf{Q} - \bf{1} + \bf{2} + \bf{3}} \Bigl\{ \Bigr.
\left[l_{\bf{1}} l_{\bf{2}} l_{\bf{3}}+ {\Gamma_{\bf{K}}}\,m_{\bf{1}} 
m_{\bf{2}} m_{\bf{3}} \right]  \ald{1} \alo{2} \alo{3} \nn
 &+&  \left[m_{\bf{1}} l_{\bf{2}} l_{\bf{3}}+{\Gamma_{\bf{K}}}\, 
m_{\bf{2}} m_{\bf{3}} l_{\bf{1}} \right] \alo{2} \alo{3} \beo{\uminus 1} +
 2\left[l_{\bf{1}} l_{\bf{2}} m_{\bf{3}}+
{\Gamma_{\bf{K}}}\, m_{\bf{2}} l_{\bf{3}} m_{\bf{1}} \right] 
 \ald{1} \alo{2} \bed{\uminus 3} 
\nn &+& 2\left[m_{\bf{1}} l_{\bf{2}} m_{\bf{3}}+{\Gamma_{\bf{K}}}\, l_{\bf{1}} 
m_{\bf{2}} l_{\bf{3}} \right]\alo{2} \bed{\uminus 3} \beo{\uminus 1} 
+  \left[l_{\bf{1}} m_{\bf{2}} m_{\bf{3}}+
{\Gamma_{\bf{K}}}\, m_{\bf{1}} l_{\bf{2}} l_{\bf{3}} \right] 
\ald{1}\bed{\uminus 2}\bed{\uminus 3}
\nn \Bigl. 
&+& \left[{m_{\bf{1}} m_{\bf{2}} m_{\bf{3}} + 
\Gamma_{\bf{K}}}l_{\bf{1}} l_{\bf{2}} l_{\bf{3}} \right] 
\bed{\uminus 2}\bed{\uminus 3}\beo{\uminus 1} \Bigr\}					
\end{eqnarray}
and
\begin{eqnarray}
	S^{-}(\bf{Q}) &=& \sqrt{2SN} \left( \left[l_{\bf{Q}} + \Gamma_{\bf{Q}}m_{\bf{Q}}\right]\ald{Q} +  
	\left[m_{\bf{Q}} + \Gamma_{\bf{Q}}l_{\bf{Q}}\right]\beo{-Q} \right) 
\end{eqnarray}
Note, that the external momentum $\bf{Q}$ can also take values outside the first MBZ and that 
$S({\bf{Q}}) \neq S({\bf{Q}+\bf{g}})$. 
Due to the non-hermiticity of the Dyson Maleev representation we have
$S({\bf{-Q}})^{-} \neq \left(S({\bf{Q}})^{+}\right)^{\dagger} $ and, thus, 
both operators have to be transformed independently.

\subsection{Continuous similarity transformation: CST}

\subsubsection{General approach}

The particular form of the flowing Hamiltonian and observables is defined 
in second quantization as 
an expansion in terms of normal-ordered operators. We write
\begin{equation}
  H(\ell) = \sum_i h_i(\ell) A_i \ \ \  O(\ell) = \sum_i o_i(\ell) D_i 
\end{equation} 
with constant monomials of bosonic operators $A_i, D_i$ and $\ell$-dependent 
coefficients $h_i(\ell), o_i(\ell)$. In this way, the flow equations turn into a 
system of differential equations for the scalar coefficients $h_i(\ell)$ and $o_i(\ell)$. 
In order to determine thesee equations, we evaluate the commutators $\left[\eta(\ell),H(\ell) \right]$ 
and $\left[\eta(\ell),O(\ell) \right]$, normal-order the results and equate the 
coefficients on both sides of the flow equation. 
In general, new operators $A_i$ and $D_i$ will occur in the commutators, which are not present in
 the initial Hamiltonian and observable. The proliferation of such operators leads 
to an infinitely large operator basis. In order to obtain a closed set of differential equations, 
a systematic and physically justified truncation scheme is required. 
Here, we use the scaling dimension of operator terms as a truncation criterion because it is particularly suitable to gapless systems, see main text and 
Ref.\  \cite{powal15}.

\subsubsection{Truncation of the Hamiltonian: Scaling dimension}

Here, we take into account the flow of operator terms up to a scaling dimension of $d=2$. 
Thus, we may neglect hexatic operator terms with $n=6$ which have scaling dimension $d=4$.
The resulting flowing Hamiltonian reads
\begin{align} 
H(\ell)& = E_0 + 
\sum_{\bf1} \w{1} \left(\ad{1}\aoo{1}+\bd{1}\bo{1}\right)+\coeffgamma{1}
\left(\ad{1}\bd{\uminus1}+\aoo{1}\bo{\uminus1}\right)
\non\\
 &+ \sum_{\bf 1,2,3,4} \Bigl\{\Coeff{1}{1} {2} {3} {4 } 
\ad{1}\ad{2}\aoo{3}\aoo{4}+			\Coeff{2}{1} {2} {3} {4 } 
\ad{1}\aoo{2}\bd{3}\bo{4} + \Coeff{3}{1} {2} {3} {4 } 
			\bd{1}\bd{2}\bo{3}\bo{4} \Bigr.
\non\\
&+ \Coeff{4}{1} {2} {3}{ 4 } \ad{1}\ad{2}\aoo{3}\bd{4}  + 
			\Coeff{5}{1} {2} {3} {4 } \ad{1}\bd{2}\bd{3}\bo{4} + 
			\Coeff{6}{1} {2} {3} {4 } \ad{1}\aoo{2}\aoo{3}\bo{4}
\non\\ 
\Bigl. &+ \Coeff{7}{1} {2} {3} {4 } \aoo{1}\bd{2}\bo{3}\bo{4} +
					\Coeff{8}{1} {2} {3} {4 } \ad{1}\ad{2}\bd{3}\bd{4} 
					+ \Coeff{9}{1} {2} {3} {4 } \aoo{1}\aoo{2}\bo{3}\bo{4} \Bigr\}. \quad
\end{align}
Then, the flowing generator $\eta(\ell)$ is  given by 
\begin{align} 
\eta(\ell) &=  
\sum_{\bf1}\coeffgamma{1}\left(\ad{1}\bd{\uminus1}-\aoo{1}\bo{\uminus1}\right)
\non\\
      &+ \sum_{\bf 1,2,3,4} \Bigl\{\Coeff{4}{1} {2} {3} {4 } 
			\ad{1}\ad{2}\aoo{3}\bd{4}  +  \Coeff{5}{1} {2} {3} {4 } 
			\ad{1}\bd{2}\bd{3}\bo{4} - \Coeff{6}{1} {2} {3} {4 } 
			\ad{1}\aoo{2}\aoo{3}\bo{4}
\non\\
\Bigl. 
&-\Coeff{7}{1} {2} {3} {4 } \aoo{1}\bd{2}\bo{3}\bo{4} +
	\Coeff{8}{1} {2} {3} {4 } \ad{1}\ad{2}\bd{3}\bd{4} - 
	\Coeff{9}{1} {2} {3} {4 } \aoo{1}\aoo{2}\bo{3}\bo{4} \Bigr\}. \quad
\end{align}

The coefficients $E_0$, $\w{1}$, $\coeffgamma{1}$, and $\Coeff{i}{1} {2} {3} {4 }$ 
depend on the flow parameter $\ell$ and satisfy the initial conditions
\begin{subequations}
\begin{eqnarray} 
  E_0\bigl|_{\ell=0}&=&-2J (S^2+AS +A^2/4)
\\
 \w{1}\bigl|_{\ell=0}&=& 2J(2S+A)\sqrt{1-\gamma_{\bf 1}^2}
\\
   \coeffgamma{1}\bigl|_{\ell=0}&=&0
\\
   \Coeff{1}{1}{2}{3}{4}\bigl|_{\ell=0}&=& -  {l_1l_2l_3l_4}
	\frac {J}{N} V_{1234}^{(1)} \dG{1+2-3-4} 
\\
   \Coeff{2}{1}{2}{3}{4}\bigl|_{\ell=0}&=& -  4{l_1l_2l_3l_4}
	\frac {J}{N}V_{1\uminus42\uminus3}^{(4)} \dG{1-2+3-4}
\\
   \Coeff{3}{1}{2}{3}{4}\bigl|_{\ell=0}&=& -  {l_1l_2l_3l_4}
	\frac {J}{N}V_{\uminus4\uminus3\uminus2\uminus1}^{(9)} \dG{1+2-3-4}      
\end{eqnarray}
\begin{eqnarray} 
   \Coeff{4}{1}{2}{3}{4}\bigl|_{\ell=0}&=& -  2{l_1l_2l_3l_4}
	\frac {J}{N}V_{1 2\uminus4 3}^{(3)} \dG{1+2-3+4}
\\
   \Coeff{5}{1}{2}{3}{4}\bigl|_{\ell=0}&=& -  2{l_1l_2l_3l_4}
	\frac {J}{N}V_{\uminus4 \uminus1\uminus2\uminus3}^{(6)} \dG{1+2+3-4}
\\
   \Coeff{6}{1}{2}{3}{4}\bigl|_{\ell=0}&=& -  2{l_1l_2l_3l_4}
	\frac {J}{N}V_{1\uminus4 2 3}^{(2)} \dG{1-2-3-4}      
\\
   \Coeff{7}{1}{2}{3}{4}\bigl|_{\ell=0}&=& -  2{l_1l_2l_3l_4}
	\frac {J}{N}V_{\uminus4\uminus3 1\uminus2}^{(5)} \dG{-1+2-3-4}
\\
   \Coeff{8}{1}{2}{3}{4}\bigl|_{\ell=0}&=& -  {l_1l_2l_3l_4}
	\frac {J}{N}V_{ 1 2\uminus3\uminus4}^{(7)} \dG{1+2+3+4}
\\
   \Coeff{9}{1}{2}{3}{4}\bigl|_{\ell=0}&=& -  {l_1l_2l_3l_4}
	\frac {J}{N}V_{\uminus3\uminus4 1 2}^{(8)} \dG{-1-2-3-4}\quad .
\end{eqnarray}
\label{eq:initialcondition}
\end{subequations}
The explicit vertex functions $V_{1234}^{(i)}$ are defined in 
Ref.~\cite{uhrig13}.

\subsubsection{Flow equations: Hamiltonian}
\label{app:flowham}

Inserting $H(\ell)$ and $\eta (\ell)$ into the flow equation 
\mbox{$\partial_{\ell} H(\ell) = \left[\eta(\ell),H(\ell) \right]$} and keeping 
all operators up to scaling dimension $d=2$, we obtain the following differential 
equations
\begin{subequations}
\begin{align}                   
\partial_{\ell} E_0 &= 
-8\sum_{ {1,2,3,4}} \Coeff{8}{1} {2} {3} {4 }
\Coeff{9}{1} {2} {3} {4} \dG{1+2+3+4} + -2\sum_{1}
\coeffgamma{1}\coeffgamma{1}
\\   
\partial_{\ell}\w{1}&=(-2)\,\coeffgamma{1}\coeffgamma{1} + 
\sum_{ {3,4}} \Bigl\{(-4)\,\Coeff{4}{3} {4} {1} {5}  
\Coeff{6}{1} {3} {4} {5}\dG{3+4-1+5}\Bigr.  
\nn
&+ (-16)\,\Coeff{8}{1} {3} {4} {5 }
\Coeff{9}{1} {3} {4} {5 }\dG{3+4+1+5} \Bigr\} 
\nn
 &+\sum_{ {3}} \Bigl\{(-4)\,\Coeff{8}{1} {3} {1} {\uminus3}\coeffgamma{3}   
-4\Coeff{6}{1} {1} {3} {\uminus3}\coeffgamma{3} \Bigr\} 
\\ 
\partial_{\ell} \coeffgamma{1}&=(-2)\,\coeffgamma{1}\w{1} + 
\sum_{{3,4,5}} \Bigl\{ (-8)\,\Coeff{4}{3}{4}{1}{5}
\Coeff{9}{3}{4}{\uminus1}{5}\dG{-1+3+4+5}\Bigr.
\nn
&+(-8)\,\Coeff{5}{3}{4}{5}{\uminus1}\Coeff{9}{1}{3}{4}{5} \dG{1+3+4+5}\Bigr\} 
\nn
&+\sum_{{3}} \Bigl\{ (-1)\,\Coeff{2}{1}{3}{\uminus1}{\uminus3}
\coeffgamma{3}+(-8)\Coeff{8}{1}{3}{\uminus1}{\uminus3}\coeffgamma{3}\Bigr\}
\end{align}
																																				
\begin{align} 
 \partial_{\ell}  \Coeff{1}{1} {2} {3} {4 }  &=\dG{1+2-3-4}\Bigl\{
(-1)\, \Coeff{4}{1}{2}{4}{\uminus3 }\coeffgamma{3} + \Bigr. 
(-1)\,\Coeff{4}{1} {2} {3} {\uminus4 }\coeffgamma{4}
\non\\
&+(-1)\, \Coeff{6}{2} {3} {4} {\uminus1 }\coeffgamma{1}+
(-1)\, \Coeff{6}{1} {3} {4} {\uminus2 }\coeffgamma{2}+\nonumber\\ 
	&\sum_{{5,6}}(-2)\, \Coeff{4}{2} {5} {4} {6 }\Coeff{6}{1} {3} {5} {6 }
	\dG{2+5-4+6}+ \nonumber\\ 
					& (-2)\, \Coeff{4}{1} {5} {3} {6 }\Coeff{6}{2} {4} {5} {6 }\dG{1+5-3+6}+ \nonumber\\
					&(-2)\, \Coeff{4}{2} {5} {3} {6 }\Coeff{6}{1} {4} {5} {6 }\dG{1+5-3+6}+ \nonumber\\ 
					&(-2)\, \Coeff{4}{1} {5} {4} {6 }\Coeff{6}{2} {3} {5} {6 }\dG{1+5-3+6}+ \nonumber\\ 
					& \Bigl. 	(-4)\, \Coeff{8}{1} {2} {5} {6 }\Coeff{9}{3} {4} {5} {6 }\dG{1+2+5+6}\Bigr\} \quad
\end{align}

\begin{align}  
 \partial_{\ell}  \Coeff{2}{1} {2} {3} {4 }  &=  \dG{1-2+3-4}\Bigl\{
(-4)\, \Coeff{4}{1} {\uminus4} {2} {3 }\coeffgamma{\uminus4} \Bigr. 
+(-4)\, \Coeff{5}{1} {3} {\uminus2} {4 }\coeffgamma{2}
\nn 	
&+(-4)\, \Coeff{6}{1} {2} {\uminus3} {4 }\coeffgamma{\uminus3}+
(-4)\, \Coeff{7}{2} {3} {4} {\uminus1 }\coeffgamma{1}+
\nn 
&\sum_{\bf{5,6}}(-4)\, \Coeff{4}{5} {6} {2} {3 }\Coeff{6}{1} {5} {6} {4 }\dG{5+6-2+3}
\nn
&+(-4)\, \Coeff{5}{1} {5} {6} {4 }\Coeff{7}{2} {3} {5} {6 }\dG{1+5+6-4} 
\nn 
&+(-8)\, \Coeff{4}{1} {5} {2} {6 }\Coeff{7}{5} {3} {4} {6 }\dG{1+5-2+6} 
\nn
&+(-8)\, \Coeff{5}{5} {3} {6} {4 }\Coeff{6}{1} {2} {5} {6 }\dG{5+3+6-4}
\nn
&+ \Bigl.  	(-32)\, \Coeff{8}{1} {5} {3} {6 }\Coeff{9}{2} {5} {4} {6 }\dG{1+5+3+6}\Bigr\} .
\end{align}

\begin{align}   
\partial_{\ell} \Coeff{3}{1} {2} {3} {4 }  &=\dG{1+2-3-4}   \Bigl\{ 
(-1)\, \Coeff{5}{\uminus3} {1} {2} {4 }\coeffgamma{\uminus3}+ \Bigr. 
(-1)\,  \Coeff{5}{\uminus4} {1} {2} {3 }\coeffgamma{\uminus4 }
\nn	
&+  (-1)\,  \Coeff{7}{\uminus2} {1} {3} {4 }\coeffgamma{\uminus2 } 
+ (-1)\,  \Coeff{7}{\uminus1} {2} {3} {4 }\coeffgamma{\uminus1 } + 
\nn
&\sum_{\bf{5}}(-4)\,  \Coeff{5}{5} {2} {6} {4 }\Coeff{7}{5} {1} {3} {6 }
\dG{5+1+6-4} +
\nn 
&  (-4)\,  \Coeff{5}{5} {1} {6} {3 }\Coeff{7}{5} {2} {4} {6 }\dG{5+1+6-3}   +
\nn
&  (-4)\,  \Coeff{5}{5} {2} {6} {3 }\Coeff{7}{5} {1} {4} {6 }\dG{5+1+6-4} +
\nn 
&  (-4)\,  \Coeff{5}{5} {1} {6} {4 }\Coeff{7}{5} {2} {3} {6 }\dG{5+1+6-3}   +
\nn
&   \Bigl. (-4)\,  \Coeff{8}{5} {6} {1} {2 }\Coeff{9}{5} {6} {3}{4 }\dG{5+6+1+2} \Bigr\}
\end{align}

\begin{align}   
\partial_{\ell} \Coeff{4}{1} {2} {3} {4 } &=\dG{1+2-3+4} \Bigl\{ 
\left(\w{3} - \w{1} - \w{2} - \w{4}\right)\, \Coeff{4}{1} {2} {3} {4 } \Bigr.	
\nn
&+  (-2)\,\Coeff{1}{1}{2}{3}{\uminus4}\coeffgamma{\uminus4}+ 
(-4)\,\Coeff{8}{1}{2}{4}{\uminus3}\coeffgamma{3} + 
\nn  
&     \left(-\frac{1}{2}\right)\,\big(\Coeff{2}{2}{3}{4}{\uminus1}
\coeffgamma{1} +\Coeff{2}{1}{3}{4}{\uminus2}\coeffgamma{2}\big)+               
\nn                   		    
&\sum_{\bf{5,6}}(-2)\, \Coeff{1}{1} {2} {5} {6 }\Coeff{4}{5} {6} {3} {4 }
\dG{1+2-5-6} +  
\nn  
&(-1)\,\Coeff{2}{2} {5} {4} {6 }\Coeff{4}{1} {5} {3} {6 }\dG{2-5+4-6} + 
\nn 
& (-1)\, \Coeff{2}{1} {5} {4} {6 }\Coeff{4}{2} {5} {3} {6 }\dG{1-5+4-6}  +  
\nn 
 & (-8)\, \Coeff{6}{2} {3} {5} {6 }\Coeff{8}{1} {5} {4} {6 }\dG{2-3-5-6}+    
\nn                                         
 &   (-8)\, \Coeff{6}{1} {3} {5} {6 }\Coeff{8}{2} {5} {4} {6 }\dG{1-3-5-6} +  
\nn 
&\Bigl. (-4)\, \Coeff{7}{3} {4} {5} {6 }\Coeff{8}{1} {2} {5} {6 }\dG{3-4+5+6} \Bigr\}  
\end{align}

\begin{align}  
    \partial_{\ell} \Coeff{5}{1} {2} {3} {4 }&= \dG{1+2+3-4} \Bigl\{ 
	\left(-\w{1} - \w{2} - \w{3} + \w{4}\right)\, \Coeff{5}{1} {2} {3} {4 }+	\Bigr. 
\nn
&  (-2)\,\Coeff{3}{2}{3}{4}{\uminus1}\coeffgamma{1} + (-4)\,
\Coeff{8}{1}{\uminus4}{2}{3}\coeffgamma{\uminus4} + 
\nn    
&     \left(-\frac{1}{2}\right)\,\left(\Coeff{2}{1}{\uminus2}{3}{4}
\coeffgamma{\uminus2} +\Coeff{2}{1}{\uminus3}{2}{4}\coeffgamma{\uminus3}\right)+
 \nn          		    
&\sum_{\bf{5,6}} (-2)\, \Coeff{3}{2} {3} {5} {6} \Coeff{5}{1} {5} {6} {4 }\dG{2+3-5-6} +    
\nn
& (-1)\, \Coeff{2}{1} {5} {3} {6 }\Coeff{5}{5} {2} {6} {4 }\dG{1-5+3-6} +    
\nn
&  (-1)\, \Coeff{2}{1} {5} {2} {6 }\Coeff{5}{5} {3} {6} {4 }\dG{1-5+2-6}  +   
\nn
&  (-8)\, \Coeff{7}{5} {3} {4} {6 }\Coeff{8}{1} {5} {2} {6 }\dG{5-3+4+6}  +   
\nn                                       
&  (-8)\, \Coeff{7}{5} {2} {4} {6 }\Coeff{8}{1} {5} {3} {6 }\dG{5-2+4+6}  +   
\nn
&  \Bigl. (-4)\, \Coeff{6}{1} {5} {6} {4 }\Coeff{8}{5} {6} {2} {3}\dG{1-5-6-4} \Bigr\}
\end{align}

\begin{align}  
\partial_{\ell} \Coeff{6}{1} {2} {3} {4 } &= \dG{1-2-3-4}\Bigl\{ 
\left(\w{1} - \w{2} - \w{3}- \w{4}\right)\, 
\Coeff{6}{1} {2} {3} {4 }	+\Bigr.	\nn  
& (-2)\,\Coeff{1}{1}{\uminus4}{2}{3}\coeffgamma{\uminus4} +
 (-4)\,\Coeff{9}{2}{3}{4}{\uminus1}\coeffgamma{1} + 
\nn
& \left(-\frac{1}{2}\right)\,\big(\Coeff{2}{1}{3}{\uminus2}{4}\coeffgamma{2} +\Coeff{2}{1}{2}{\uminus3}{4}\coeffgamma{3}\big) +
\nn        			        		    
 &\sum_{\bf{5,6}} (-2)\, \Coeff{1}{5} {6} {2} {3 }\Coeff{6}{1} {5} {6} {4 }
\dG{5+6-2-3}  +  
\nn 
&  (-1)\, \Coeff{2}{5} {3} {6} {4 }\Coeff{6}{1} {2} {5} {6 }\dG{5-3+6-4} +
\nn
 &  (-1)\, \Coeff{2}{5} {2} {6} {4 }\Coeff{6} {1} {3} {5} {6 }\dG{5-2+6-4} +
\nn
  & (-8)\, \Coeff{4}{1} {5} {3} {6 }\Coeff{9}{2} {5} {4} {6 }\dG{1+5-3+6} +
 \nn 
 &      (-8)\, \Coeff{4}{1} {5} {2} {6 }\Coeff{9}{3} {5} {4} {6 }\dG{1+5-2+6} + 
\nn  
 &  \Bigl. (-4)\, \Coeff{5}{1} {5} {6} {4 }\Coeff{9}{2} {3} {5} {6 }\dG{1+5+6-4}  \Bigr\}
\end{align}

\begin{align}  
\partial_{\ell} \Coeff{7}{1} {2} {3} {4 } &= \dG{1-2+3+4} \Bigl\{ 
\left(-\w{1} + \w{2} - \w{3} - \w{4}\right)\, \Coeff{7}{1} {2} {3} {4 }+\Bigr. 
\nn 
 &  (-2)\,\Coeff{3}{\uminus1}{2}{3}{4}\coeffgamma{1} +
 (-4)\,\Coeff{9}{1}{\uminus2}{3}{4}\coeffgamma{2} + 
\nn    
&         \left(-\frac{1}{2}\right)\,\big(\Coeff{2}{\uminus4}{1}{2}{3}\coeffgamma{4} +\Coeff{2}{\uminus3}{1}{2}{4}\coeffgamma{3}\big) +
\nn	   		     		    
&\sum_{\bf{5,6}} (-2)\,\Coeff{3}{5} {6} {3} {4 }\Coeff{7}{1} {2} {5} {6 }
\dG{5+6-3-4} +  
\nn  
&  (-1)\,\Coeff{2}{5} {1} {6} {4 }\Coeff{7}{5} {2} {3} {6 }\dG{5-1+6-4} +
\nn
& (-1)\,\Coeff{2}{5} {1} {6} {3 }\Coeff{7}{5} {2} {4} {6 }\dG{5-1+6-3} +
\nn
&(-8)\,\Coeff{5}{5} {2} {6} {4 }\Coeff{9}{1} {5} {3} {6 }\dG{5+2+6-4}   + 
\nn                                        
& (-8)\,\Coeff{5}{5} {2} {6} {3 }\Coeff{9}{1} {5} {4} {6 }\dG{5+2+6-3}  + 
\nn  
&  \Bigl.(-4)\,\Coeff{4}{5} {6} {1} {2 }\Coeff{9}{5} {6} {3} {4 }\dG{5+6-1+2}  \Bigr\}
\end{align}

\begin{align}     
   \partial_{\ell} \Coeff{8}{1} {2} {3} {4 } &= \dG{1+2+3+4}\Bigl\{ 
\left(-\w{1} - \w{2} - \w{3} - \w{4}\right)\, \Coeff{8}{1} {2} {3} {4 }\Bigr.+ 
	\nn			  
& \sum_{\bf{5,6}} (-2)\, \Coeff{1}{1} {2} {5} {6 }
\Coeff{8}{5} {6} {3} {4 }\dG{1+2-5-6}+ 
\nn
& (-1)\, \Coeff{2}{2} {5} {4} {6 }\Coeff{8}{1} {5} {3} {6 }\dG{2-5+4-6} +
\nn
&(-1)\,\Coeff{2}{1} {5} {4} {6 }\Coeff{8}{2} {5} {3} {6 }\dG{1-5+4-6} +
\nn
&(-1)\, \Coeff{2}{2} {5} {3} {6 }\Coeff{8}{1} {5} {4} {6 }\dG{2-5+3-6} +       
\nn                                    
&(-1)\, \Coeff{2}{1} {5} {3} {6 }\Coeff{8}{2} {5} {4} {6 }\dG{1-5+3-6} +    
\nn 
&\Bigl. (-2)\, \Coeff{3}{3} {4} {5 }{6 }\Coeff{8}{1} {2} {5} {6 }\dG{3+4-5-6}  \Bigr\}
\end{align}

\begin{align} 
\partial_{\ell} \Coeff{9}{1} {2} {3} {4 }  &= \dG{1+2+3+4} \Bigl\{
\left(-\w{1} - \w{2} - \w{3} - \w{4}\right) \Coeff{9}{1} {2} {3} {4 }+\Bigr.
\nn		     		    
&\sum_{ \bf{5,6}} (-2) \, \Coeff{1}{5} {6} {1} {2 }\Coeff{9}{5} {6} {3} {4 }\dG{5+6-1-2}   + 
\nn 
& (-1)\, \Coeff{2}{5} {2} {6} {4 }\Coeff{9}{1} {5} {3} {6 }\dG{5-2+6-4} +
\nn
&(-1)\,\Coeff{2}{5} {1} {6} {4 }\Coeff{9}{2} {5} {3} {6 }\dG{5-1+6-4} +
\nn
&(-1) \,\Coeff{2}{5} {2} {6} {3 }\Coeff{9}{1} {5} {4} {6 }\dG{5-2+6-3}   +  
\nn                                       
&(-1) \,\Coeff{2}{5} {1} {6} {3 }\Coeff{9}{2} {5} {4} {6 }\dG{5-1+6-3} +   
\nn 
&\Bigl.   (-2)\, \Coeff{3}{5} {6} {3} {4 }\Coeff{9}{1} {2} {5} {6 }\dG{3+4-5-6}  \Bigr\}
\end{align}
\end{subequations}

\subsubsection{Truncation of observables}

Similar to the Hamilton operator, the flowing observables have to be truncated as well. 
But the corresponding spin operators are not evaluated directly, but as part of a resolvent. 
Hence, we do not truncate the spin operators in terms of scaling dimension. Instead, we keep those terms 
which couple the ground state to the relevant magnon channels. As we are interested in the subspaces with 
up to three magnons, we include operator terms up to a cubic level in annihilation and creation operators.
Then, the flowing observables read
\begin{equation}
S^{z}({\bf{Q}}) = \sum_{\bf{1,2}}  \Bigl\{  			
\szI{Q}{1}{2}   \ald{1}\alo{2} +\szII{Q}{1}{2}    \bed{1}\beo{2}	
+\szIII{Q}{1}{2}  \alo{1}\beo{2}    +\szIV{Q}{1}{2}   \ald{1}\bed{2}
\Bigr\} + C(\bf{Q})													
\end{equation}
with initial conditions 
\begin{subequations}
\begin{eqnarray}
 \szI{Q}{1}{2}\vert_{\ell=0}  &=& \dG{\bf{Q}-\bf{1}+\bf{2}}\, 
\left[{\Gamma_{\bf{Q-1+2}}} l_{\bf{1}} l_{\bf{2}}-\,m_{\bf{1}} m_{\bf{2}} \right] \pp
 \szII{Q}{1}{2}\vert_{\ell=0}   &=& \dG{\bf{Q}-\bf{1}+\bf{2}}\, 
  \left[{\Gamma_{\bf{Q-1+2}}} m_{\bf{1}} m_{\bf{2}}-\,l_{\bf{1}} l_{\bf{2}} \right]\pp
 \szIII{Q}{1}{2}\vert_{\ell=0}     &=&\dG{\bf{Q}+\bf{1}+\bf{2}}\, 
  \left[{\Gamma_{\bf{Q-1+2}}} m_{\bf{1}} l_{\bf{2}}-\,l_{\bf{1}} m_{\bf{2}} \right]\pp
 \szIV{Q}{1}{2}\vert_{\ell=0}   &=&\dG{\bf{Q}-\bf{1}-\bf{2}}\, 
 \left[{\Gamma_{\bf{Q-1+2}}} m_{\bf{1}} l_{\bf{2}}-\,m_{\bf{1}} l_{\bf{2}} \right]\pp
C(\bf{Q})\vert_{\ell=0} &=&	\left( m_{Q}^2 -S N\right)\left(\Gamma_{\bf{Q}}-1 \right)\dG{\bf{Q}}								
\end{eqnarray}
\end{subequations}
,
\begin{align}
S^{+}({\bf{Q}}) &= \sum_{\bf{1}} \spI{Q}{1} \,\alo{1} \, + \, \spII{Q}{1} \,  \bed{1} +\sum_{\bf{1,2,3}}   \Bigl\{  \Bigr.  \spIII{Q}{1}{2}{3}\, 
	\bed{1}\bed{2}\beo{3}\nn
	&+\,\spIV{Q}{1}{2}{3}\,  \ald{1}\alo{2}\alo{3}
	+\,\spV{Q}{1}{2}{3}\,   \ald{1}\bed{2}\bed{3} 
	+\,\spVI{Q}{1}{2}{3}\,  \alo{1}\bed{2}\beo{3}
	\nn 
	&+\,\spVII{Q}{1}{2}{3}\, \ald{1}\alo{2}\bed{3} +\,\spVIII{Q}{1}{2}{3}\,\alo{1}\alo{2}\beo{3} \Bigl. \Bigr\}			
\end{align}
with initial conditions 
\begin{subequations} 
\begin{eqnarray}
 \spI{Q}{1}\vert_{\ell=0}           &=&  \dG{\bf{Q+1}}\sqrt{2SN} 
\left( 1 -  \frac{1}{S N} \sum_{\bf{k}} m_{\bf{k}}^2 \right) 
\left(l_{\bf{1}} + {\Gamma_{\bf{1}}} \, m_{\bf{1}}\right)                   
\\
 \spII{Q}{1}\vert_{\ell=0}            &=&  \dG{\bf{Q-1}}\sqrt{2SN} 
\left( 1 -  \frac{1}{S N} \sum_{\bf{k}} m_{\bf{k}}^2 \right)                      
\left(m_{\bf{1}} + {\Gamma_{\bf{1}}} \, l_{\bf{1}}\right)
\\
 \spIII{Q}{1}{2}{3}\vert_{\ell=0}     &=&
\dG{\bf{Q}-\bf{1}-\bf{2}+\bf{3}}\, \left[m_{\bf{1}} m_{\bf{2}} m_{\bf{3}}+ 
{\Gamma_{\bf{Q-1-2+3}}}\,l_{\bf{1}} l_{\bf{2}} l_{\bf{3}} \right]\pp
 \spIV{Q}{1}{2}{3}\vert_{\ell=0}      &=&\dG{\bf{Q}-\bf{1}+\bf{2}+\bf{3}}\, 
\left[l_{\bf{1}} l_{\bf{2}} l_{\bf{3}}+{\Gamma_{\bf{Q-1+2+3}}}\,m_{\bf{1}} 
m_{\bf{2}} m_{\bf{3}} \right]
\pp
 \spV{Q}{1}{2}{3}\vert_{\ell=0}       &=&
\dG{\bf{Q}-\bf{1}-\bf{2}-\bf{3}}\, \left[l_{\bf{1}} m_{\bf{2}} m_{\bf{3}}
+{\Gamma_{\bf{Q-1-2-3}}}\,m_{\bf{1}} m_{\bf{2}} l_{\bf{3}} \right]
\pp
 \spVI{Q}{1}{2}{3}\vert_{\ell=0}      &=&         2 \,\dG{\bf{Q}+\bf{1}-\bf{2}+
\bf{3}}\, \left[l_{\bf{1}} l_{\bf{2}} l_{\bf{3}}+{\Gamma_{\bf{Q+1-2+3}}}\,
m_{\bf{1}} m_{\bf{2}} m_{\bf{3}} \right]\pp
 \spVII{Q}{1}{2}{3}\vert_{\ell=0}     &=&         2 \,\dG{\bf{Q}-\bf{1}+\bf{2}-
\bf{3}}\, \left[l_{\bf{1}} l_{\bf{2}} l_{\bf{3}}+{\Gamma_{\bf{Q-1+2-3}}}\,
m_{\bf{1}} m_{\bf{2}} m_{\bf{3}} \right]\pp
 \spVIII{Q}{1}{2}{3}\vert_{\ell=0}    &=&\dG{\bf{Q}+\bf{1}+\bf{2}+\bf{3}}\, 
\left[l_{\bf{1}} l_{\bf{2}} l_{\bf{3}}+{\Gamma_{\bf{Q+1+2+3}}}\,m_{\bf{1}} 
m_{\bf{2}} m_{\bf{3}} \right] 								
\end{eqnarray}
\end{subequations}
and 
\begin{align}
S^{-}({\bf{Q}}) &=  \smI{Q} \,  \ald{Q} \, + \, \smII{-Q} \,  \beo{-Q} 
               + \sum_{\bf{1,2,3}} \Bigl\{  \Bigr. 
							\smIII{Q}{1}{2}{3}\, \ald{1}\ald{2}\bed{3} \nn
					&		+\,\smIV{Q}{1}{2}{3}\,  \ald{1}\ald{2}\alo{3} 
					+\,\smV{Q}{1}{2}{3}\,   \ald{1}\bed{2}\beo{3}     
					+\,\smVI{Q}{1}{2}{3}\,  \ald{1}\alo{2}\beo{3} 
					\nn
					& +\,\smVII{Q}{1}{2}{3}\, \bed{1}\beo{2}\beo{3}     
					+\,\smVIII{Q}{1}{2}{3}\,\alo{1}\beo{2}\beo{3}	\Bigl.  \Bigr\}	
					\label{eq:inital_sm}		 										
\end{align}
with initial conditions 
\begin{subequations}
\begin{align}
 \smI{Q}\vert_{\ell=0}  &=  \sqrt{2SN}  
\left(l_{\bf{Q}} + {\Gamma_{\bf{Q}}} \, m_{\bf{Q}}\right)\pp
 \smII{Q}\vert_{\ell=0} &=  \dG{\bf{Q-1}}\sqrt{2SN}                       
\left(m_{\bf{Q}} + {\Gamma_{\bf{Q}}} \, l_{\bf{Q}}\right)\pp
 \smIII{Q}{1}{2}{3}\vert_{\ell=0}     &=         0 \qquad
 \smIV{Q}{1}{2}{3}\vert_{\ell=0}      =         0 \qquad
 \smV{Q}{1}{2}{3}\vert_{\ell=0}       =         0 \pp
 \smVI{Q}{1}{2}{3}\vert_{\ell=0}      &=         0 \qquad
 \smVII{Q}{1}{2}{3}\vert_{\ell=0}     =         0 \qquad
 \smVIII{Q}{1}{2}{3}\vert_{\ell=0}    =        0			 \quad.
\end{align}
\end{subequations}
The flow of $C(\Q)$ does not influence the other coefficients of the observables 
and it is not required in the evaluation of the relevant resolvents. 
Thus, it is not considered in the flow equations.

\subsubsection{Flow equations: Observables}
\label{app:flowobs}

The flow equations of the observables read
\begin{subequations}
 \begin{align} 
 \partial_{\ell}  \szI{Q} {1} {2 }  &=\dG{Q+1-2}\Bigl\{ \Bigr.  
	(-1) \coeffgamma{1}\szIII{Q} {2} {\uminus 1}+	(-1) \coeffgamma{2}\szIV{Q}{1} 
	{\uminus 2}
					\nn 
		&+ \sum_{{3,4}} (-2) \Coeff{4}{1} {3} {2} {4 }\szIII{Q} {3} {4}	\dG{1+3-2+4}+ 
					\nn					 
					&\Bigl. \qquad (-2) \Coeff{6}{1} {2} {3} {4 }\szIV{Q} {3} {4}\dG{1-2-3-4}\Bigr\} 
\end{align}
 \begin{align}   
 \partial_{\ell}  \szII{Q} {1} {2 } &= \dG{Q+1-2}\Bigl\{ \Bigr.  
(-1) \coeffgamma{1}\szIV{Q} {\uminus2} {1}+	(-1) \coeffgamma{2}\szIII{Q}
 {\uminus1} {2}
					\nn
	& +\sum_{{3,4}} (-2) \Coeff{5}{3} {1} {4} {2 }\szIII{Q}{3}{4}\dG{3+1+4-2}+
	\nn 					 
					& \Bigl. \qquad	(-2) \Coeff{7}{3}{1}{2}{4 }\szIV{Q}{3}{4}\dG{3-1+2+4}\Bigr\} 
\end{align}
 \begin{align}
 \partial_{\ell}  \szIII{Q} {1} {2 } &= \dG{{Q-1-2}}\Bigl\{ \Bigr.  
(-1) \coeffgamma{\uminus 2}\szI{Q} {\uminus2} {1}+(-1) \coeffgamma{1}\szII{Q}
 {\uminus1} {2}
					\nn 
	&+ \sum_{{3,4}} (-4) \Coeff{9}{1} {3} {2} {4 }
	\szIV{Q}{3}{4}\dG{1+3+2+4}\Bigr\}
	\end{align}
 \begin{align}  
 \partial_{\ell}  \szIV{Q} {1} {2 }  &= \dG{Q+1+2}\Bigl\{ \Bigr.  
(-1) \coeffgamma{\uminus 2}\szI{Q} {1} {\uminus2}+
(-1) \coeffgamma{1}\szII{Q} {2} {\uminus1}
					\nn
					&+\sum_{{3,4}} (-4) \Coeff{8}{1} {3} {2} {4 }\szIV{Q}{3}{4}\dG{1+3+2+4}\Bigr\} 					 					 
\end{align}
\end{subequations}
\begin{subequations}
  \begin{align}   
 \partial_{\ell}  \spI{Q}  &=   (-1)\coeffgamma{Q}\spII{\uminus Q}+ 
\sum_{{2}}\Bigl\{     (-1)\coeffgamma{2}\spVII{Q} {2} {Q} {\uminus 2}+ 
   (-2)\coeffgamma{2}\spVIII{Q} {Q} {2} {\uminus 2}\Bigr\}
					\nn 	 
					&+ \sum_{{2,3,4}} \Bigl\{(-2) \Coeff{4}{2} {3} {Q} {4 }
					\spVIII{Q} {2} {3} {4}\dG{2+3-Q+4}\Bigr. 
					\nn 					 
					& + \Bigl. 	   (-4) 
					\Coeff{9}{Q} {2} {3} {4 }\spV{Q} {2} {3} {4}+\dG{Q+2+3+4}\Bigr\}  
\end{align}
 \begin{align}   
 \partial_{\ell}  \spII{Q}  &=  -\coeffgamma{\uminus Q}\spI{\uminus Q}+ 
	\sum_{{2}}\Bigl\{-2\coeffgamma{2}\spV{\uminus} {Q} {2} {Q} {\uminus 2}-\coeffgamma{2}\spVI{\uminus Q} {2} {Q} {\uminus 2} \Bigr\}
					\nn 	 
					&+ \sum_{{2,3,4}} \Bigl\{(-2) \Coeff{7}{2} {Q} {3} {4 }
					\spV{\uminus Q} {2} {3} {4}\dG{2-Q+3+4}\Bigr. \nn 					 
					&+ \Bigl. (-4) \Coeff{8}{2} {3} {Q} {4 }
					\spVIII{\uminus Q} {2} {3} {4}\dG{Q+2+3+4} \Bigr\} 
\end{align}
   \begin{align}  
 \partial_{\ell}  \spIII{Q} {1} {2} {3}  &= \dG{Q+1+2-3}\Bigl\{ \Bigr. 
-\Coeff{5}{Q} {1} {2} {3}\spI{Q}-\coeffgamma{\uminus 3}\spV{\uminus Q}
 {\uminus3} {1} {2} 
					\nn 	
	&+	\left(-\frac{1}{2}\right)\coeffgamma{\uminus 1}\spVI{Q} {\uminus1} {2} {3}+  \left(-\frac{1}{2}\right)\coeffgamma{\uminus 2}\spVI{ Q} {\uminus2} {1} {3}  
					\nn
					&+ \sum_{{4,5}} (-2) \Coeff{7}{4} {1} {3} {5 }\spV{Q} {4} {2} {5}
					\dG{4-1+3-5} +  
					\nn 
					& (-2) \Coeff{7}{4} {1} {2} {5 }\spV{Q} {4} {3} {5}\dG{4-1+2-5}+  
					\nn  
					&(-1) \Coeff{5}{4} {1} {5} {3 }\spVI{Q} {4} {1} {5}\dG{4+1+5-3} +  
					\nn
					& (-1) \Coeff{5}{4} {2} {5} {3 }\spVI{Q} {4} {2} {5}\dG{4+2+5-3}+   
					\nn
					& \Bigl.   (-2) \Coeff{8}{4} {5} {1} {2 }\spVIII{Q} {4} {5} {3}
					\dG{4+5+1+2} \Bigr\} 
\end{align}
   \begin{align}   
 \partial_{\ell}  \spIV{Q} {1} {2} {3} &=\dG{Q+1-2-3}\Bigl\{ \Bigr. -
\Coeff{6}{1}{2}{3}{\uminus Q}\spII{\uminus Q}-
\coeffgamma{1}\spVIII{Q}{2}{3}{\uminus1}
\nn & + 
\left(-\frac{1}{2}\right)\coeffgamma{2}\spVII{Q}{1}{3}{\uminus2}+
\left(-\frac{1}{2}\right)\coeffgamma{3}\spVII{Q}{1}{2}{\uminus3}+ 
					\nn
					& \sum_{{4,5}} (-2) \Coeff{4}{1}{4}{2}{5 }
					\spVIII{Q}{3}{4}{5}\dG{1+4-2+5}+   
					\nn 
					&
					(-2) \Coeff{4}{1}{4}{3}{5 }\spVIII{Q}{2}{4}{5}\dG{1+4-3+5}+   
					\nn  
					&(-2) \Coeff{9}{2}{3}{4}{5 }\spV{Q}{1}{4}{5}\dG{2+3+4+5}+   
					\nn
					&(-1) \Coeff{6}{1}{2}{4}{5 }\spVII{Q}{4}{3}{5}\dG{1-2-4-5}+  
					\nn
					&(-1) \Coeff{6}{1}{3}{4}{5 }\spVII{Q}{4}{2}{5}\dG{1-3-4-5}\Bigr\}  
 \end{align}
 \begin{align} 
 \partial_{\ell}  \spV{Q}{1}{2}{3}  &= \dG{Q+1+2+3}\Bigl\{ \Bigr.  
\Coeff{5}{1}{2}{3}{\uminus Q}\spII{\uminus Q} -2
\Coeff{8}{1}{Q}{2}{3}\spI{Q}
					\nn 
					&-\coeffgamma{1}\spIII{ Q}{2}{3}{\uminus1}-\left(\frac{1}{2}\right)
					[\coeffgamma{\uminus 2}\spVII{Q}{1}{\uminus2}{3}+  
					\coeffgamma{\uminus 3}\spVII{Q}{1}{\uminus3}{2}]  
					\nn
					&+\sum_{{4,5}}(-2) \Coeff{8}{4}{5}{2}{3}\spIV{Q}{1}{4}{5}\dG{4+5+2+3}+   
					\nn 
					& (-2) \Coeff{8}{1}{4}{2}{5}\spVI{Q}{4}{3}{5}\dG{1+4+2+5}+   
					\nn  
					&\Bigl.    (-2) \Coeff{8}{1}{4}{3}{5}\spVI{Q}{4}{2}{5}\dG{1+4+3+5}\Bigr\}
  \end{align}
 \begin{align}
 \partial_{\ell}  \spVI{Q}{1}{2}{3}  &=\dG{Q-1+2-3}\Bigl\{ \Bigr.  
-2\Coeff{7}{1}{2}{3}{\uminus Q}\spII{\uminus Q}-2
\coeffgamma{\uminus2}\spVIII{Q}{1}{\uminus2}{3}
\nn &+ (-2)\coeffgamma{1}\spIII{Q}{2}{\uminus1}{3}+ (-1)\coeffgamma{\uminus3}\spVII{Q}{\uminus3}{1}{2}+  
					\nn &
					\sum_{{4,5}}(-8) \Coeff{9}{1}{4}{3}{5}\spV{Q}{4}{2}{5}\dG{1+4+3+5}+   
					\nn
					& (-4) \Coeff{5}{1}{4}{2}{5}\spVIII{Q}{1}{4}{5}\dG{1+4+2-5}+   
					\nn
					&(-2) \Coeff{7}{1}{2}{4}{5}\spIII{Q}{4}{5}{3}\dG{1-2+4+5}+   
					\nn
					&(-2) \Coeff{7}{4}{2}{3}{5}\spVII{Q}{4}{1}{5}\dG{4-2+3+5}+   
					\nn
					&\Bigl.   (-2) \Coeff{4}{4}{5}{1}{2}\spVIII{Q}{4}{5}{3}\dG{4+5-1+2}\Bigr\}  
\end{align}
\begin{align} 
 \partial_{\ell}  \spVII{Q}{1}{2}{3} & = \dG{Q+1-2+3}\Bigl\{ \Bigr.  
-2\Coeff{4}{1}{Q}{2}{3}\spI{Q}-2\coeffgamma{\uminus3}\spIV{Q}{1}{2}{\uminus3}
\nn &
					-2\coeffgamma{2}\spV{Q}{1}{3}{\uminus2}+(-1)\coeffgamma{1}
					\spVI{Q}{2}{3}{\uminus1} 
					\nn &+
					\sum_{{4,5}}(-8)\Coeff{8}{1}{4}{3}{5 }
					\spVIII{Q}{2}{4}{5}\dG{1+4+3+5} +  
					\nn 
					&
					(-4)\Coeff{6}{1}{2}{4}{5 }\spV{Q}{4}{3}{5}\dG{1-2-4-5}+   
					\nn 
					& (-2)\Coeff{4}{4}{5}{3}{3 }\spIV{Q}{1}{4}{5}\dG{4+5-3+3}+   
					\nn
					& (-2)\Coeff{7}{2}{3}{4}{5 }\spV{Q}{1}{4}{5}\dG{2-3+4+5}+   
					\nn
					& \Bigl.   (-2)\Coeff{4}{1}{4}{2}{5 }\spVI{Q}{4}{3}{5}\dG{1+4-2+5}\Bigr\}  
 \end{align}
   \begin{align}  
 \partial_{\ell}  \spVIII{Q}{1}{2}{3}  & =\dG{Q-1-2-3}\Bigl\{ \Bigr.  
\Coeff{6}{Q}{1}{2}{3}\spI{Q}-2\Coeff{9}{1}{2}{3}{\uminus Q}\spII{\uminus Q}
\nn &  -\coeffgamma{\uminus 3}\spIV{Q}{\uminus3}{1}{2}- 
\left(\frac{1}{2}\right)[\coeffgamma{1}\spVI{Q}{2}{\uminus1}{3}+  
					\coeffgamma{2}\spVI{Q}{1}{\uminus2}{3}] 
					\nn & +
					\sum_{{4,5}}(-2) \Coeff{9}{1}{2}{4}{5 }\spIII{Q}{4}{5}{3}\dG{1+2+4+5}+  
					\nn
					&(-2) \Coeff{9}{1}{4}{3}{5 }\spVII{Q}{4}{2}{5}\dG{1+4+3+5}+   
					\nn
					& \Bigl.    (-2) \Coeff{9}{2}{4}{3}{5 }
					\spVII{Q}{4}{1}{5}\dG{2+4+3+5}\Bigr\}  
 \end{align}
\end{subequations}
\begin{subequations}
   \begin{align}  
 \partial_{\ell}  \smI{Q}  &=   -\coeffgamma{Q}\smII{\uminus Q}+  
	\Bigl\{ -\coeffgamma{2}\smVI{\uminus Q} {Q} {2} {\uminus 2}+ \Bigr. 
						\Bigl.	-2\coeffgamma{2}\smIII{\uminus Q} {Q} {2} {\uminus 2} \Bigr\}
						\nn & +
					\sum_{{2,3,4}}	\Bigl\{(-2) \coeff{6}{Q 2 3 4 }\smIII{\uminus Q} {2} {3} {4} \dG{Q-2-3-4}\Bigr.  
					\nn 					 
					& +\Bigl.  (-4) \coeff{8}{Q 2 3 4 }\smVIII{Q} {2} {3} {4}\dG{Q-2+3+4}\Bigr\}  
\end{align}
 \begin{align}  
 \partial_{\ell}  \smII{Q} &=   -\coeffgamma{Q}\smI{Q}-2
\coeffgamma{2}\smVIII{Q} {2} {Q} {\uminus 2}-
\coeffgamma{2}\smV{ Q} {2} {\uminus2} {Q} \Bigr\}
\nn & + 	\sum_{{2,3,4}}
					\Bigl\{(-2) \coeff{5}{2 3 4 Q }\smVIII{Q}{2}{3}{4}\dG{2+3+4-Q} \Bigr.  
					\nn					 
					& +\Bigl. 	   (-4) \coeff{9}{2 3 Q 4 }\smIII{Q} {2} {3} {4}\dG{Q+2+3+4} \Bigr\} 
  \end{align}
 \begin{align}   
 \partial_{\ell}  \smIII{Q} {1} {2} {3} &= \dG{Q+1+2+3}\Bigl\{ \Bigr.  
\Coeff{4}{1}{2}{\uminus Q}{3}\smI{\uminus Q}   
-\coeffgamma{\uminus 3}\smIV{Q} {1} {2} {Q}
\nn & - \left(\frac{1}{2}\right)[
\coeffgamma{1}\smV{Q} {2} {3} {\uminus1}+
\coeffgamma{2}\smV{ Q} {1} {3} {\uminus2}]-2\Coeff{8}{1}{2}{3}{Q}\smII{Q} 
					\nn & + 
					\sum_{{4,5}}
					(-2) \Coeff{8}{1}{2}{4}{5}\smVII{Q} {3} {4} {5}\dG{1+2+4+5}   
					\nn 
					&+(-2) \Coeff{8}{1}{4}{3}{5}\smVI{Q} {2} {4} {5}\dG{1+3+4+5}   
					\nn
					& +\Bigl.  (-2) \Coeff{8}{2}{4}{3}{5}\smVI{Q} {1} {4} {5}\dG{2+3+4+5}\Bigr\} 
  \end{align}
 \begin{align}  
 \partial_{\ell}  \smIV{Q} {1} {2} {3}  &= \dG{Q+1+2-3}\Bigl\{ \Bigr. -
\Coeff{5}{1}{2}{3}{Q}\smII{Q}-\coeffgamma{3}\smIII{ Q }{1} {2} {\uminus3}  
					\nn 					
					&-	\left(\frac{1}{2}\right) [ \coeffgamma{1}\smVI{Q} {2} {3} {\uminus1}+  \coeffgamma{2}\smVI{ Q} {1} {3} {\uminus2}]   
					\nn 					&+
					\sum_{{4,5}}(-1) \Coeff{4}{1}{4}{3}{5}\smVI{Q} {2} {4} {5}\dG{1+4-3+5}+  
					\nn 
					& (-1) \Coeff{4}{2}{4}{3}{5}\smVI{Q} {1} {4} {5}\dG{2+4-3+5}+  
					\nn 
					&(-2) \Coeff{6}{1}{3}{4}{5}\smIII{Q} {2} {4 }{5}\dG{1-3-4-5}+   
					\nn
					&(-2) \Coeff{6}{2}{3}{4}{5 }\smIII{Q }{1}{4} {5}\dG{2-3-4-5} +  
					\nn
					&\Bigl. 	   (-2) \Coeff{8}{1}{2}{4}{5}\smVIII{Q }{3} {4} {5}
					\dG{1+2+4+5}\Bigr\} 
\end{align}
   \begin{align}  
 \partial_{\ell}  \smV{Q} {1} {2} {3}  & =\dG{Q+1+2-3}\Bigl\{ \Bigr.  
-2\Coeff{5}{1}{2}{Q}{3}\smII{Q}-2\coeffgamma{\uminus 3}\smIII{Q}{1} 
{\uminus3} {2}					
\nn &    (-1)\coeffgamma{\uminus 2}\smVI{Q} {1} {\uminus2} {3}+  
					 (-2)\coeffgamma{1}\smVII{ Q} {2} {3} {\uminus1}
					\nn
					& + \sum_{{4,5}}   (-2) \Coeff{6}{1}{4}{5}{3}\smIII{Q} {4} {5} {2}
					\dG{1-4-5-3}  +
					\nn &  
					(-4) \Coeff{7}{4}{2}{3}{5}\smIII{Q} {1} {4} {5}\dG{-4+2-3-5}    +     
					\nn
					& (-2) \Coeff{5}{4}{2}{5}{3}\smVI{Q} {1} {4} {5}\dG{4+2+5-3}  +       
					\nn
					&(-2) \Coeff{5}{1}{4}{5}{3}\smVII{Q} {2} {4} {5}\dG{1+4+5-3} +        
					\nn
					& \Bigl.  (-8) \Coeff{8}{1}{4}{2}{5}\smVIII{Q} {4} {3} {5}\dG{1+4+2+5}\Bigr\}  
 \end{align}
   \begin{align}  
 \partial_{\ell}  \smVI{Q} {1} {2} {3} &= \dG{Q+1+2-3}\Bigl\{ \Bigr.-2\Coeff{6}{1}{2}{\uminus Q}{3}\smI{\uminus Q}-2\coeffgamma{\uminus3}\smIV{ Q} {1} 
{\uminus3} {2} 
					\nn
					&+(-2)\coeffgamma{1}\smVIII{ Q} {2} {3} {\uminus1}+  
					(-1)\coeffgamma{2}\smV{ Q} {1} {\uminus2} {3}  
					\nn & +
					\sum_{{4,5}} (-2) \Coeff{6}{1}{4}{5}{3}\smIV{Q} {4} {5} {2}\dG{1-4-5-3}+ 
					\nn 
					&(-2) \Coeff{6}{1}{2}{4}{5}\smV{Q} {4} {5} {3}\dG{1-2-4-5}+   
					\nn 
					&(-2) \Coeff{5}{1}{4}{5}{3}\smVIII{Q} {2} {4} {5}\dG{1+4+5-3}+   
					\nn
					&(-4) \Coeff{4}{1}{4}{2}{5}\smVIII{Q} {4} {3} {5}\dG{1+4-2+5}+   
					\nn
					&\Bigl.(-8) \Coeff{9}{2}{4}{3}{5}\smIII{Q} {1} {4} {5}\dG{2+4+3+5}\Bigr\} 
	\end{align}
   \begin{align} 
 \partial_{\ell}  \smVII{Q} {1} {2} {3}  &= \dG{Q+1-2-3}\Bigl\{\Bigr. 
-\Coeff{8}{\uminus Q}{1}{2}{3}\smI{\uminus Q} 
					-\coeffgamma{\uminus1}\smVIII{ Q} {\uminus1} {2} {3}  
					\nn 
					& -	 \left(\frac{1}{2}\right)[\coeffgamma{\uminus2}\smV{ Q} {\uminus2}
					{1} {3}+  \coeffgamma{\uminus3}\smV{ Q} {\uminus3} {1} {2}]  
					\nn &+ 
					\sum_{{4,5}}
						(-1)\Coeff{7}{4}{1}{2}{5}\smV{Q} {4} {5} {3}\dG{4-1+2+5}+   
					\nn 
					& (-1)\Coeff{7}{4}{1}{3}{5}\smV{Q} {4} {5} {2}\dG{4-1+3+5}+   
					\nn 
					& (-2)\Coeff{5}{4}{1}{5}{2}\smVIII{Q} {4} {3} {5}\dG{4+1-5+2}+   
					\nn
					& (-2)\Coeff{5}{4}{1}{5}{3}\smVIII{Q} {4} {2} {5}\dG{4+1-5+3}+   
					\nn
					&\Bigl.(-2)\Coeff{9}{4}{5}{2}{3}\smIII{Q} {4} {5} {1}\dG{2+4+3+5}\Bigr\} 
\end{align}
  \begin{align}  
 \partial_{\ell}  \smVIII{Q}{1}{2}{3} &= \dG{Q-1-2-3}\Bigl\{\Bigr.  
\Coeff{7}{1}{Q}{2}{3}\smII{Q}
					-2\Coeff{9}{1}{\uminus Q}{2}{3}\smI{\uminus Q}  -
					\nn &
					\coeffgamma{1}\smVII{Q}{\uminus1}{2}{3}-  
					 \left(\frac{1}{2}\right)[\coeffgamma{\uminus2}\smVI{Q}{\uminus2}{1}{3}+  \coeffgamma{\uminus3}\smVI{Q}{\uminus3}{1}{2}]  
					\nn & +
					\sum_{{4,5}} (-2) \Coeff{9}{4}{5}{2}{3}\smIV{Q}{4}{5}{1}\dG{2+4+3+5}+   
					\nn
					& (-2) \Coeff{9}{1}{4}{2}{5} \smV{Q}{4}{5}{3}\dG{1+4+2+5}+   
					\nn
					&\Bigl. (-2) \Coeff{9}{1}{4}{3}{5}\smV{Q}{4}{5}{2}\dG{1+4+3+5} \Bigr\}  .
\end{align}
\end{subequations}

\section{Spectral densities}
\label{app:B}

The general description of spectral densities is given in the main text in Sect.\ II.B. Explicit 
formulae are given in the sequel.

\subsection{Transversal dynamic structure factor}

The transversal dynamic structure factor is given by
\begin{eqnarray}
   S^{xx+yy}(\omega, \Q)  &=& -\frac{1}{\pi} \Im{\left(\bra{0}S^{-}_{\textnormal{eff}}(-\Q) \frac{1}{\omega  
	- (H_{\textnormal{eff}}-  \bar{E}_0)} S^{+}_{\textnormal{eff}}(\Q)\ket{0} \right)}              
\end{eqnarray}
It splits into a one-magnon contribution
\begin{subequations}
\begin{eqnarray}
   S^{xx+yy}(\omega, \Q)\left|_{1\textnormal{mag}}\right.  &=&
	                           -\frac{1}{\pi} \Im{\left(\bra{0}\smII{-Q} \beo{Q} \frac{1}{\omega  - (H_{\textnormal{eff}}-  \bar{E}_0)} 
														\spII{Q} \bed{Q} \ket{0}\right)} \\
			&=&  \smII{-Q}\spII{Q} \delta \left(\omega- \omega(\Q)\right)
\end{eqnarray}
\end{subequations}
where $W^{1\textnormal{mag}} = \smII{-Q}\spII{Q}$ defines the one-magnon spectral weight, and a three-magnon contribution
\begin{align}
   &S^{xx+yy}(\omega, \Q)\left|_{3\textnormal{mag}}\right. =
	\nn & \qquad  \sum_{\bf{1},\bf{2}} \bra{0}\alo{2}\beo{Q-2} \szIII{-Q}{2}{Q-2} 
	\frac{1}{\omega  - (H_{\textnormal{eff}}-  \bar{E}_0)} \szIV{Q}{1}{Q-1} \ald{1}\bed{Q-1}\ket{0} 
	\end{align}
which constitutes the incoherent part of the spectral density. The transversal static structure factor 
$W^{3-\textnormal{mag}}(\bf{Q})  = \sum_{\bf{1}}\szIII{-Q}{1}{Q-1} \szIV{Q}{1}{Q-1} $ 
provides the total weight in the spectral density 
$S^{xx+yy}(\omega, \Q)\left|_{3\textnormal{mag}} \right.$,
i.e., in the three-magnon channel.
 The static structure factor in the thermodynamic limit 
is obtained by an extrapolation in $\frac{1}{L}\to 0$.

\subsection{Longitudinal dynamic structure factor}

The longitudinal dynamic structure factor is given by
\begin{subequations}
\begin{align}
   &S^{zz}(\omega, \Q)  = \bra{0}S^z_{\textnormal{eff}}(-\Q) \frac{1}{\omega  - 
	(H_{\textnormal{eff}}-  \bar{E}_0)} 
	S^z_{\textnormal{eff}}(\Q)\ket{0} \\
& \qquad = \sum_{\bf{1},\bf{2}} \bra{0}\alo{2}\beo{Q-2} \szIII{-Q}{2}{Q-2} 
\frac{1}{\omega  - (H_{\textnormal{eff}}-  \bar{E}_0)} \szIV{Q}{1}{Q-1} \ald{1}\bed{Q-1}\ket{0}
\end{align}
\end{subequations}
Accordingly, its total weight is determined by the longitudinal static structure factor 
\begin{equation}
W^{2-\textnormal{mag}}(\bf{Q}) = \sum_{\bf{1}\bf{2}}\szIII{-Q}{1}{Q-1} \szIV{Q}{1}{Q-1}
\end{equation}
The static structure factor in the thermodynamic limit is obtained by a linear extrapolation 
in $\frac{1}{L}\to 0$.

\subsection{Spectral weights}
\label{app:specweights}

Integrating the spectral densities over frequency yields the spectral weights. 
They can be classified according
to the number $n$ of magnons. The transversal weight is denoted by 
$S^{xx}_n(\Q)+S^{yy}_n(\Q)$ and the
longitudinal weight by $S^{zz}_n(\Q)$. The extrapolated values 
for the thermodynamic limit are given in Tab.\ \ref{tab:weights}

\begin{table}[ht]
\begin{center}
\begin{tabular}{|c|c||l|l|l||l|}
\hline 
$Q_x$            & $Q_y           $ & $S^{xx}_1+S^{yy}_1$ & $S^{xx}_3+S^{yy}_3$ & relative  & $S^{zz}_2$\\
\hline\hline
$\frac{\pi}{2} $ & $\frac{\pi}{2} $ & 0.5839  & 0.1337 & 0.8137 & 0.2508 \\ \hline
$\frac{\pi}{3} $ & $\frac{2\pi}{3}$ & 0.5456  & 0.1749 & 0.7573 & 0.2496 \\ \hline
$\frac{\pi}{6} $ & $\frac{5\pi}{6}$ & 0.4705  & 0.2557 & 0.6479 & 0.2470 \\ \hline
$0             $ & $\pi           $ & 0.4339  & 0.2952 & 0.5951 & 0.2457 \\ \hline
$\frac{\pi}{3} $ & $\pi           $ & 0.6575  & 0.2788 & 0.7022 & 0.3234 \\ \hline
$\frac{2\pi}{3}$ & $\pi           $ & 1.5146  & 0.3313 & 0.8205 & 0.6177 \\ \hline
$\pi           $ & $\pi           $ & -       & -      & -      & 1.1063 \\ \hline
$\frac{5\pi}{6}$ & $\frac{5\pi}{6}$ & 2.1826  & 0.3508 & 0.8615 & 0.8002 \\ \hline
$\frac{2\pi}{3}$ & $\frac{2\pi}{3}$ & 0.9576  & 0.2759 & 0.7763 & 0.4314 \\ \hline
$\frac{\pi}{2} $ & $\frac{\pi}{2} $ & 0.5839  & 0.1337 & 0.8137 & 0.2508 \\ \hline
$\frac{\pi}{3} $ & $\frac{\pi}{3} $ & 0.3435  & 0.0620 & 0.8471 & 0.1300 \\ \hline
$\frac{\pi}{6} $ & $\frac{\pi}{6} $ & 0.1603  & 0.0190 & 0.8941 & 0.0402 \\ \hline
$0             $ & $0             $ & 0       & 0      & 0      & 0      \\ \hline
$0             $ & $\frac{\pi}{3} $ & 0.2235  & 0.0351 & 0.8643 & 0.0707 \\ \hline
$0             $ & $\frac{2\pi}{3}$ & 0.3997  & 0.1570 & 0.7180 & 0.1853 \\ \hline
$0             $ & $\pi           $ & 0.4339  & 0.2952 & 0.5951 & 0.2457 \\
\hline
\end{tabular}
\end{center}
\caption{
\label{tab:weights}
Spectral weights extrapolated to infinite system size for various channels at various high-symmetry points in the MBZ.
The transversal channel is comprises the one-magnon and the three-magnon channel. 
The fourth column shows the relative weight of the one-magnon channel, i.e., the ratio
$(S^{xx}_1+S^{yy}_1)/(S^{xx}_1+S^{yy}_1+ S^{xx}_3+S^{yy}_3)$.
At $\Q=(0,0)$ the DSF are identical zero because the corresponding operator does not
introduce any excitation. At $\Q=(\pi,\pi)$ the static transversal DSF diverges so that no value can be given.
The last column displays the two-magnon weight in the longitudinal channel.
The lattice constant is set to unity. 
}
\end{table}

\subsection{Non-symmetric continued fraction representation}

The spectral densities in the multi-magnon subspaces are 
evaluated using a continued fraction representation for the resolvent \cite{viswa94}.
Here, we are dealing with a non-symmetric problem
\begin{equation}
  R(\omega) = \bra{v_L} \frac{1}{\omega-H}\ket{v_R}
\end{equation}
with $H\neq H^{\dagger}$ and $\bra{v_L} \neq \bra{v_R}$ which leads to a generalized continued fraction representation
\begin{equation}
   R(\omega) = \cfrac{\braket{v_L}{v_R}}{ a_0 - \omega - \cfrac{b_1 g_1}{a_1 - 
	\omega -\cfrac{b_2 g_2}{\cdots}} }
\end{equation}
where we omitted any further indices on the magnon sector or the momentum for the sake of clarity.
The coefficients $a_i$, $b_i$ and $g_i$ can be obtained by a non-symmetric Lanczos tridiagonalization \cite{rajak91} 
for $H$ with starting vectors $\bra{v_L}$ and $\ket{v_R}$ yielding the tridiagonal matrix
\begin{eqnarray}
H_{\textnormal{tri}} \ \ = \ \ 
\begin{bmatrix}
a_0    & b_1     & 0      &\cdots     & 0      \\
g_1    & a_1     & b_2    &\ddots     & \vdots       \\
0      & g_2     & a_2    &\ddots     & 0       \\
\vdots & \ddots  & \ddots &\ddots     & b_{m}\\
0   	 & \cdots  &   0    & g_{m}   & a_m  \phantom{\vdots}
\end{bmatrix}
\end{eqnarray}
where $m$ is the number of Lanczos steps. The corresponding spectral density is a 
sum of weighted $\delta$-peaks located at the eigenvalues $\omega_i$ of $H_{\textnormal{tri}}$
\begin{equation}
 I(\omega) = -\frac{1}{\pi} \Im{\,R(\omega)} 
           = \sum_{i=0}^n W_i \delta(\omega - \omega_i) .
\end{equation}
which becomes a continuous function in the limit $m\rightarrow \infty$.

The weights $W_i$ are given by the overlap between the starting vectors and 
the corresponding left and right eigenvector of $H_{\textnormal{tri}}$, i.e.,
$W_i = \braket{L_i}{v_R} \braket{v_L}{R_i}$ with 
$\bra{L_i}H=  \bra{L_i}\omega_i$ and $H\ket{R_i}$=$\omega_i\ket{R_i}$.
Note, that the eigenbasis is bi-orthonormal, meaning that $\braket{L_i}{R_i}=\delta_{ij}$.

The asymptotic behavior of coefficients is related to the upper bound $E_u$ and 
lower bound $E_l$ of the continuum
\begin{subequations}
\begin{eqnarray}
    E_u &=& a_\infty + 2 \sqrt{g_{\infty}b_{\infty}} \\
		E_l &=& a_\infty - 2 \sqrt{g_{\infty}b_{\infty}},
\end{eqnarray}
\end{subequations}
see for instance Ref.\ \cite{petti85}.
Hence, it is possible to approximate the coefficients $a_m$ and $g_m b_m$ by  
$a_{\infty}$ and $g_{\infty} b_{\infty}$ for sufficiently large $m$ in the thermodynamic limit.
In the case of a finite Hilbert space the behavior of coefficients will deviate from this asymptotic rule at larger 
$m$ depending on the actual size of the Hilbert space. To keep these finite-size effects minimum, we
enlarge the Hilbert space by appropriate interpolation schemes increasing the number of grid points in the MBZ.
This increases the maximum number $m_{\textnormal{max}}$ of Lanczos steps before the coefficients start to 
deviate sizeably from the asymptotic behavior.
For $m>m_{\textnormal{max}}$ the sequence of coefficients is approximated by constant coefficients 
$a_{\infty}$ and $g_{\infty}b_{\infty}$. In generic recursion approaches, this infinite sequence in the 
continued fraction representation is equivalent to appropriate terminating functions 
\cite{viswa94,petti85}.

In order to reproduce the finite resolution in the experiment, we replace the $\delta$-peaks resulting
from the diagonalization of the finite tridiagonal matrix $H_\text{tri}$
by Gaussian distribution functions with the corresponding weight $W_i$ and an 
artificial broadening which fits the resolution of the experimental data
  \begin{eqnarray}
	\sum_i W_i \delta(\omega - \omega_i) \rightarrow I(\omega) = \sum_i W_i  \frac{1}{\sqrt{2 \sigma^2 \pi}} \textnormal{e}^{ \frac{1}{2}\left(\frac{\omega -\omega_i}{\sigma}\right)^2}\quad.
	 \end{eqnarray}
The values of $W_i$ and $\omega_i$ are found from the diagonalization of 
the interpolated system and the overlap with its eigen vectors. We normalize the broadened 
spectral functions $(\sum_i W_i)^{-1}I(\omega)$ and rescale them by the corresponding
spectral weights in the thermodynamic limit. 
The latter are obtained by an extrapolation of the static structure factors in $1/L\to 0$, see 
Tab.\ \ref{tab:weights}. Thus, the continua in Fig.\ 4 in the main text are finally plotted according to 
\begin{subequations}
\begin{eqnarray}
I^{xx+yy}_\text{plot}(\omega) &=& \frac{S^{xx}_3(\Q)+S^{yy}_3(\Q)}{\sum_i W_i^{xx+yy}} 
I^{xx+yy}(\omega)
\\
I^{zz}_\text{plot}(\omega) &=& \frac{S^{zz}_2(\Q)}{\sum_i W_i^{zz}} I^{zz}(\omega)
\end{eqnarray}
\end{subequations}
in the transversal and in the longitudinal channel, respectively.
This way, we avoid any bias of the relative total weights due to different 
interpolation schemes applied for different magnon sectors.

\subsection{Interpolation schemes}

\subsubsection{Dispersion}

For the dispersion $\omega(\k)$ we are using a Lanczos re-sampling method 
\cite{lancz38}  which is suitable for the interpolation of periodic functions. 
In order to treat the linear cusp at  $\k=0$ appropriately we interpolate the 
function $f(\k) = \omega(\k)^2$ instead, taking the square root afterwards.

\subsubsection{Observables}

The coefficient functions in the longitudinal observable depend on a two-dimensional wavevector like the dispersion. 
But these functions exhibit poles at certain momenta which makes it more complicated to interpolate them directly. 
In order to overcome this problem, we interpolate the function $f(\k) = 
\frac{\bar{s}(\k)}{s(\k)+c}$ where 
$\bar{s}(\k)$ is the function transformed by the CST and $s(\k)$ denotes the corresponding function \emph{before} 
the CST which is known analytically. As the renormalization of the CST does not shift or alter the poles in 
$\bar{s}(\k)$ the function $f(\k)$ is finite and smooth. 
The constant shift $c$ is introduced to avoid zeroes in the denominator. 
This requires to track sign changes in $s(\k)$ 
due to umklapp processes induced by the corresponding factors $\Gamma(\bf{K})$. 
Typically, we set the shift to $c=\alpha\cdot \textnormal{min}(s(k))$ with $\alpha\approx 1.5$

The coefficient functions in the three-magnon contribution of the transversal observable are directly interpolated using a 
quadrilinear interpolation scheme. Note, that the interpolation scheme used in the longitudinal part cannot be applied here because
the functions  in (\ref{eq:inital_sm}) before the CST are identical to zero. We stress that the three-magnon response is \emph{only} induced
by the renormalization in the course of the CST. 
The interpolation for the three-magnon contributions does not need to be as sophisticated as 
for the two-magnon contribution because the required number of 
intermediate grid points is much smaller compared to the longitudinal case. 
This is so because at fixed total momentum $\Q$ the dimension of 
the Hilbert space in the two-magnon
subsector is $L^2$ while it is $L^4$ in the three-magnon sector.

\subsubsection{Interaction}

The vertex functions with their six-dimensional arguments are interpolated directly by a multilinear 
or nearest neighbor interpolation scheme depending on the required number of grid points. 
The arguments of the vertex functions $C_i(\k_1,\k_2,\k_3,\k_4)$ are momenta defined in the first MBZ. 
Thus, the reciprocal vector $\Gamma = \sum_i k_i$ determined by the total momentum conservation may switch 
inside the first MBZ leading to a sign change in the coefficients. 
In order to avoid such jumps we track sign changes in the interpolation scheme.

\subsection{General procedure}

In the following, we summarize the general procedure to calculate the continuous spectral densities 
briefly:
\begin{enumerate}
	\item 
	First, we perform a linear extrapolation of the effective coefficients at $L=8$ and $L=16$ in 
	$\frac{1}{L}$ to $L=\infty$. We obtain an extrapolated effective model defined at 
	$N=8 \times 8$ grid points in the first MBZ. 
	The same extrapolation scheme is applied to the static structure factors yielding the total weights of the  two- and three-magnon contributions. 
	\item 
	We interpolate the effective coefficients in order to increase the Hilbert space for 
	the subsequent recursion analysis to reduce finite-size effects
	For the longitudinal densities, the system size is enhanced from $L=8$ to $L=192$. 
	For the transversal three-magnon continuum, a system size of $L=16$ is sufficient. 
	\item A non-symmetric Lanczos algorithm is applied to determine the continued fraction representation 
	of the spectral densities. The list of the resulting coefficients is extended further by 
	additional coefficients known from the asymptotic limit. We approximate the spectral densities by a 
	sum of Gaussian distribution functions with uniform broadening. 
	Their locations and weights are obtained by an exact diagonalization of the extended tridiagonal matrix. 
\end{enumerate}

\end{appendix}

\end{document}